\documentclass[prd,twocolumn,superscriptaddress,floatfix,amsmath,amssymb,amsfonts,longbibliography]{revtex4-2}

\usepackage{float} 
\usepackage{comment}
\usepackage[normalem]{ulem}
\usepackage[english]{babel}
\usepackage{graphicx}
\usepackage{dcolumn}
\usepackage{blindtext}
\usepackage{verbatim}
\usepackage{mathrsfs}
\usepackage{musicography}
\usepackage{amsmath}
\usepackage{cancel}
\usepackage{physics}
\usepackage{epstopdf}
\usepackage{mathtools}
\usepackage{tensor}
\usepackage{xcolor}
\usepackage[usenames,dvipsnames]{pstricks}
\usepackage{epsfig}
\usepackage{pst-grad} 
\usepackage{pst-plot} 
\usepackage{hyperref}
\usepackage{verbatim}
\usepackage{slashed}
\usepackage{dsfont}
\usepackage{lipsum}

\newcommand{\mf}{\mathsf}

\newcommand{\ii}{\mathrm{i}}

\definecolor{ForestGreen}{RGB}{0,90,0}

\renewcommand{\bm}[1]{\boldsymbol{#1}}
\newcommand{\tc}[1]{\textsc{#1}}

\newcommand{\secl}[1]{%
\section{#1}
}

\allowdisplaybreaks[1] 

\begin{document}

\title{Finite-size clocks in quantum field theory and the twin paradox}

\author{Matheus H. Zambianco}
\email{mhzambia@uwaterloo.ca}
\affiliation{Department of Applied Mathematics, University of Waterloo, Waterloo, Ontario, N2L 3G1, Canada}
\affiliation{Perimeter Institute for Theoretical Physics, Waterloo, Ontario, N2L 2Y5, Canada}
\affiliation{Institute for Quantum Computing, University of Waterloo, Waterloo, Ontario, N2L 3G1, Canada}

\author{T. Rick Perche}
\email{rick.perche@su.se}
\affiliation{Nordita, KTH Royal Institute of Technology and Stockholm University, Hannes Alfv\'{e}ns v\"{a}g 12, 23, SE-106 91 Stockholm, Sweden}

\begin{abstract}

Vacuum fluctuations in quantum field theory impose fundamental limitations on our ability to measure time at arbitrarily short scales. To investigate the impact of universal quantum field theory effects on observer-dependent time measurements, we introduce a clock model based on the vacuum decay probability of a finite-sized quantum system. This model defines an effective notion of proper time that depends on the microscopic properties of the clock and on how it samples vacuum fluctuations along its trajectory. We show that, in the long-time regime, this notion of time reduces to the usual proper time of special relativity. However, by studying a microscopic twin-paradox scenario, we find that, in general, time is not determined solely by the trajectory connecting two events, but also by how vacuum fluctuations interact with the internal structure of the clocks.

\end{abstract}

\maketitle

\secl{Introduction}Relativity predicts that the passage of time is relative: two clocks may yield two different measurements of time between the same two events, depending on their trajectory in spacetime. Perhaps the most clear example of the relativity of the passage of time is illustrated in the \emph{twin paradox}~\cite{SpacetimePhysics,TwinCylinder, TwinsKerr, TwinsSchwarzschild}, where two twins follow different worldlines---one twin remains on Earth, whereas the other travels aboard a spacecraft. Upon reunion, they find that the traveling twin ages less. The apparent paradox arises from the incorrect assumption that the twins’ kinematics are symmetric, when, in fact, only one twin accelerates throughout their motion. The age difference is a consequence of the fact that inertial trajectories locally maximize proper time between events. 

Although the relative passage of time has been verified by experiments~\cite{OldSchoolMuonsSR, science_relativity_clocks, TestTimeDilationPRL2014, Science2010Clocks, PhysRevLett.118.221102}, the classical relativistic description is not expected to hold in arbitrarily small spacetime scales. Indeed, quantum theory imposes fundamental restrictions on the precision of space and time measurements~\cite{SaleckerWigner, NG_1995}. Moreover, there is no universal representation for a time observable within quantum mechanics~\cite{pauli1980general, unruh_wald_1989, muga2008time}. Instead, time is more precisely investigated through the physical realization of clocks: systems whose dynamics can be used to parametrize the occurrence of physical events, enabling an operational measurement of elapsed time \cite{AutonomousQC2017, PhysRevApplied.20.034038, Smith2020}. Indeed, the most precise measurements of time intervals are realized by atomic clocks~\cite{atomic_clocks_review, Nicholson2015, McGrew2018}, which can measure times of the order of $10^{-18}\text{s}$~\cite{Oelker2019, PRL_single_atom_clock2019}.

In this paper, we are concerned with the observer-dependent passage of time in scales where the notion of ideal clocks from classical relativity cannot be implemented, which we study by introducing a clock model based on the vacuum decay probability of a finite-size quantum system. To illustrate the main features of this model, we use it to study a microscopic twin paradox scenario. Although a quantum version of the twin paradox with clocks in superposition was explored in~\cite{QuantumTwinsScience2019}, here we focus on the standard twin paradox scenario in scales where universal quantum field theory (QFT) effects become relevant for tracking the passage of time. Indeed, QFT yields the most accurate predictions for relativistic effects in quantum physics. It becomes particularly relevant for the description of non-inertial systems, such as in the Unruh effect \cite{Unruh1976, matsasUnruh, Takagi, LoukoUnruh} and short-time effects~\cite{L_Sriramkumar_1996, antiUnruh, Garay_2016,quantClass}. Thus, a complete study of the relativity of time at the smallest scales must take QFT into account.

Indeed, it is precisely in time-scales comparable to the system's light crossing time that the vacuum fluctuations of quantum fields plays more significant roles~\cite{quantClass}. As we will see, in a microscopic twin-paradox scenario where the clock's size is non-negligible compared to the measured times, parameters that are usually irrelevant for time measurements, such as the clock's shape and specific internal dynamics, end up playing crucial roles.

\secl{Finite-sized ideal clocks}\label{sec:clocks_ideal}
In special relativity, the proper time between events $\mf x_i$ and $\mf x_f$, as experienced by an observer undergoing a timelike curve $\mf z (s)$, is defined as
\begin{equation}
    \Delta \tau = \int_{s_{i}}^{s_{f}}\dd s \sqrt{-g_{\mu \nu}(\mf z(s)) \frac{d z^{\mu}}{ds}\frac{d z^{\nu}}{ds}},
\end{equation}
where $\mf x_{i} = \mf z(s_{i})$ and $\mf x_{f} = \mf z(s_{f})$. An ideal clock is any dynamical system which, when undergoing the trajectory $\mf z (s)$, measures this proper time. However, if the clock is to be implemented by a realistic physical system, it must have a finite spatial extension. This will be essential when we discuss clock models in QFT.

To model a finite-sized ideal clock ($C_{\text{ideal}}$), consider a Fermi Normal coordinate (FNC) system \cite{poisson} $(\tau, \bm \xi)$ around the timelike trajectory $\mf z(\tau)$, with $\tau$ the proper time of the curve. In those coordinates, $\mf z(\tau) = (\tau, \bm 0)$, and the neighbouring trajectories are $ z_{\bm \xi} = (\tau, \bm \xi)$. We introduce a spatial profile function $F(\bm \xi)\geq 0$ to account for the spatial extension of the clock. For simplicity, we restrict our attention to Minkowski spacetime, and we consider $F$ to be normalized at each hypersurface of constant $\tau$: $\int \dd^3\bm{\xi}\, F(\bm{\xi}) = 1.$
The time tracked by this finite-size ideal clock between events $\mf x_{i}$ and $\mf x_{f}$ is then
\begin{equation}
\mathcal{T}(C_{\text{ideal}}) \equiv \int_{\tau_i}^{\tau_f} \dd \tau \int \dd^3 \bm \xi \, F(\bm \xi) \ell(\tau, \bm \xi),
    \label{eq:smeared_clock_classical}
\end{equation}
where the proper time density $\ell(\tau, \bm \xi)$ is given by
\begin{equation}
    \ell (\tau, \boldsymbol{\xi}) = \sqrt{-g_{\mu \nu}(\mf z_{\boldsymbol{\xi}}(\tau)) \dv{z_{\bm \xi}^{\mu}}{\tau}\dv{z_{\bm \xi}^{\nu}}{\tau}} = \sqrt{-g_{00}(\tau, \bm \xi)}.
    \label{eq:local_proper_time_classical}
\end{equation}
This construction averages the proper time experienced by each curve of constant $\bm \xi$, with weight defined by $F(\bm \xi)$. The proper time density $\ell(\tau,\bm \xi)$ assigns to each point of the clock the infinitesimal proper time elapsed along the corresponding neighbouring trajectory. For instance, for an inertial clock we have $\ell_\tau(\tau,\bm \xi) = 1$ and for a uniformly accelerated clock with constant proper shape, we have $\ell_\tau(\tau, \bm \xi) = (1 + \bm a \cdot\bm \xi)$, where $\bm a$ is the clock's proper acceleration.

\secl{Non-periodic clocks in QFT}The most precise clocks available today are atomic clocks, whose timekeeping mechanism relies on coherent evolution between atomic energy levels \cite{atomic_clocks_review}. Nevertheless, these clocks are not naturally adapted to microscopic twin-paradox scenarios involving short interaction times and large accelerations. Their operation generally takes place over time scales that may be comparable to, or longer than, the relativistic effect one aims to resolve. In addition, the act of locally comparing or reading out two atomic clocks can disturb their coherent evolution and may require a subsequent reinitialization stage~\cite{PRL_single_atom_clock2019}. Time comparisons also typically involve external synchronization protocols, which complicate the interpretation of the clock reading as arising solely from the proper time accumulated along the clock's worldline.

To address these issues, we consider a non-periodic, decay-based clock model that does not require the preservation of a coherent periodic motion throughout its evolution. Instead, the clock is initialized in an excited state, allowed to interact locally with an external field along its worldline, and read out through its de-excitation probability. For a fixed interaction model and trajectory, this probability is governed by the decay rate accumulated between the initial and final events. The clock reading is therefore encoded in a local transition probability rather than in the phase of an externally interrogated oscillator, as in the case of atomic clocks \cite{PRL_single_atom_clock2019, atomic_clocks_review}.

We model the clock ($C$) as a finite-sized localized quantum system whose relevant internal dynamics can be effectively described by two energy levels, denoted by $\ket{g}$ and $\ket{e}$. The clock interacts locally with a free scalar quantum field, which we take to be initially prepared in the vacuum state. Its centre follows a prescribed worldline $\mf z(\tau)$, where $\tau$ is the proper time along the trajectory. The clock is initially prepared in the excited state $\ket{e}$, and the interaction with the field is switched on at the event $\mf x_i=\mf z(\tau_i)$ and switched off at the later event $\mf x_f=\mf z(\tau_f)$, both defined locally at the centre of $C$. At the end of the interaction, the probability that $C$ has de-excited to $\ket{g}$ is determined by the field correlations sampled by the localized system.

To leading order in the coupling parameter of the interaction ($\lambda$), the de-excitation probability depends only on a few parameters that determine the interaction, such as the shape and spacetime trajectory~\cite{Schlicht, Jorma, TalesBrunoEdu2020} of the clock, as well as the energy gap $\Omega$ between the states $\ket{g}$ and $\ket{e}$. Concretely, when the interaction between the clock and the quantum field is linear, its de-excitation probability (to leading order in $\lambda$), can be written as
\begin{equation}\label{eq:P}
    P_\tc{d} = \lambda^2\int \dd V \dd V' \Lambda(\mf x) \Lambda(\mf x')e^{\ii \Omega(\tau(\mf x)-\tau(\mf x'))} W(\mf x, \mf x'),
\end{equation}
where $\tau(\mf x)$ is a local extension of the clock's proper time around $\mf z (\tau)$~\cite{Schlicht,eduardo,TalesBrunoEdu2020,generalPD}, $\dd V = \sqrt{-g} \, \dd^4 \mf x$ is the invariant spacetime volume element, $W(\mf x, \mf x')$ is the vacuum two-point function of the external free quantum field, and $\Lambda(\mf x)$  determines the clock's spatial profile and the switching of the interaction. In particular, a rigid clock with FNCs $\mf x (\tau, \boldsymbol{\xi})$~\cite{TalesBrunoEdu2020,EduTalesBruno2021,generalPD} is such that
\begin{equation}\label{eq:Lambda}
    \Lambda(\mf x) = F(\boldsymbol{\xi})\mathds{1}_{[\tau_i,\tau_f]}(\tau),
\end{equation}
 where the clock tracks the elapsed time between the events $\mf z(\tau_i) = \mf x_i$ and $\mf z(\tau_f) = \mf x_f$. The function $F(\boldsymbol{\xi})$ is determined by the relevant spatial profile of the interaction between the clock and the field, which is typically fixed by the wavefunctions of the internal states $\ket{g}$ and $\ket{e}$~\cite{Unruh-Wald, generalPD, TalesJoseBrunoEdu2024, CasimirAlhambraAchimEdu, Pozas2016}. Moreover, the behaviour encoded by Eq.~\eqref{eq:P} is universal for localized quantum systems interacting with external fields. It appears, for example, in short-time light-matter interactions~\cite{Pozas2016,richard,RuhiEduRick2025}, in the interaction of fermions and neutrinos with quantum fields~\cite{neutrinos,carol}, and in larger-scale systems, such as systems interacting with Bose-Einstein condensates~\cite{unruhExp,oberthaler}. {Thus, for the purposes of this work, the clocks may be regarded as atom-like localized systems whose angular-momentum degrees of freedom are neglected~\cite{richard}.}

Assuming spacetime profiles as in Eq.~\eqref{eq:Lambda}, the total interaction time as measured by an ideal clock following the trajectory $\mf z(\tau)$ is simply $T_0 \equiv \tau_{f} - \tau_{i}$. To define the clock's asymptotic deexcitation rate, we introduce a one-parameter family of trajectories whose proper time $T$ can be taken arbitrarily large. Let $\Delta \mf x = \mf x_{f} - \mf x_{i}$ denote the separation vector between the endpoints of the original trajectory $\mf z (\tau)$. For $T > T_{0}$, define
\begin{equation}
z_{T}^\mu(\tau)
=
x_i^\mu
+
S(T)^\mu{}_\nu
\bigl[z^\nu(\tau)-x_i^\nu\bigr],
\end{equation}
where
\begin{align}
    S(T)^\mu{}_\nu
& =
\delta^\mu{}_\nu+
\left[\alpha\!\left(\frac{T}{T_0}\right)-1\right] \frac{\Delta x^\mu\Delta x_\nu}
{\Delta x^\beta\Delta x_\beta},
\end{align}
Notice that $S(T)$ leaves unchanged the components orthogonal to $\Delta\mf x$ while rescaling the component parallel to $\Delta\mf x$ by $\alpha(T/T_0)$. The function $\alpha$ is chosen to be monotonic, with $\alpha(1)=1$, and such that the total proper time of $\mf z_T$ is $T$. Reparametrizing this curve in terms of its proper time $\tau_T$ yields the trajectory $\mf z_T(\tau_T)$. It begins at $\mf x_i$ and ends at
$\mf x_i+\alpha\!\left(\frac{T}{T_0}\right)\Delta\mf x.$
The asymptotic rate is obtained by taking the limit $T\to\infty$. The clock's asymptotic de-excitation rate is defined as~\cite{Unruh1976,Unruh-Wald,DeWitt}:
\begin{equation}\label{eq:rate}
    \mathcal{F} \equiv \lim_{T \to \infty}\frac{P_{\tc{d}}(T)}{T},
\end{equation}
where $P_{\tc{d}}(T)$ denotes the de-excitation probability of the rescaled clock undergoing the trajectory $\mf z_T(\tau_T)$. When the clock's light-crossing time is negligible compared with the duration of the interaction, the corresponding de-excitation rate is finite (see, e.g., Ref.~\cite{Good2020}). This rate is determined by the clock's energy gap $\Omega$, its proper spatial profile $F(\boldsymbol{\xi})$, and, in general, by its trajectory.
{For an inertial clock in the Minkowski vacuum, the asymptotic excitation rate vanishes: apart from finite-time switching transients, energy conservation forbids the transition $\ket{g}\to\ket{e}$. By contrast, for uniformly accelerated motion the field induces a nonzero excitation rate at $\mathcal{O}(\lambda^2)$. In the stationary uniformly accelerated case, the upward and downward rates obey detailed balance at the Unruh temperature, driving the two-level system towards the corresponding thermal state~\cite{Unruh1976,Unruh-Wald}.}

{This late-time equilibration lies beyond the single-transition, leading-order regime considered here. Since the clock is initially prepared in $\ket{e}$, a re-excitation can affect its final ground-state probability only after a prior de-excitation and hence first contributes at $\mathcal{O}(\lambda^4)$, at fixed $T$. Thus, provided the de-excitation probability remains small, the leading-order response in Eq.~\eqref{eq:P} is well approximated by $P_{\tc{d}}(T)\simeq \mathcal{F}T$, so the clock mimics the behaviour of an ideal clock.}

In Appendix~\ref{sec:inertial_clocks}, we show that this linear regime is reached already at finite times in representative examples. For inertial clocks in Minkowski spacetime, the approximation $P_{\tc{d}}\simeq \mathcal{F}T$ becomes accurate for interaction times that are only moderately larger than the clock's light-crossing time and the scale set by $\Omega$. For instance, when $T \gtrsim 30\sigma$, with $\sigma$ the clock's light-crossing time, the relative deviation
$|P_{\tc{d}}-\mathcal{F}T|/(\mathcal{F}T)$
remains below $2\%$. We also verify numerically that the same behaviour persists for non-inertial trajectories within this linear-response window: as the total interaction time $T$ increases, the time registered by $C$ approaches the trajectory's proper time.
Motivated by the discussion above, we define the time registered by the clock $C$ between the events $\mf x_i$ and $\mf x_f$ as~\footnote{Notice that it is not possible to directly access the de-excitation probability using a single clock through a one-shot experiment. Instead, one should repeat the experiment $N$ times under the exact same circumstances. Then, if the clock decays in $N_{d}$ of those experiments, the elapsed time between events $\mf x_i$ and $\mf x_f$ as measured by this clock model is given by $\mathcal{T}(C) \equiv \mathcal{F}^{-1} N_{d}/N\approx  \mathcal{F}^{-1}P_\tc{d}.$}
\begin{equation}
    \mathcal{T}(C) \equiv \frac{\lambda^2}{\mathcal{F}}\int \dd V \dd V'F(\bm \xi) F(\bm \xi') \chi(\tau)\chi(\tau') e^{\ii \Omega(\tau - \tau')} W(\mf x, \mf x'),
    \label{eq:clock_explicit}
\end{equation}
with $\chi(\tau) = \mathds{1}_{[\tau_i,\tau_f]}(\tau)$.  This definition is such that, whenever the linear response regime $P_{\tc{d}}/ \mathcal{F}\approx T$ applies, $\mathcal{T}(C)$ coincides with the elapsed proper time along the reference trajectory $\mf z(\tau)$. It is also possible to rewrite this definition in the form of Eq.~\eqref{eq:smeared_clock_classical},
\begin{equation}\label{eq:newtime}
    \mathcal{T}(C) = \int_{\tau_i}^{\tau_f} \dd \tau \int \dd^3 \bm \xi \, F(\bm \xi) \ell_{\Omega}(\tau, \bm \xi),
\end{equation}
with the corresponding local proper time density
\begin{align}
    \ell_\Omega(\mf x) &  = \frac{\lambda^2}{\mathcal{F}}\sqrt{-g_{00}}\int_{\tau_i}^{\tau_f} \dd \tau'\int \dd^3 \bm \xi'\, \sqrt{-g_{00}'} F(\bm \xi') \nonumber \\ & \quad\quad\quad\times  \Re\bigg(e^{\ii \Omega(\tau(\mf x) - \tau')}W(\mf z_{\bm \xi}(\tau),\mf z_{\bm \xi'}(\tau'))\bigg).
    \label{eq:local_proper_time_QFT}
\end{align}
Notice that although the proper time prescribed by Eq.~\eqref{eq:newtime} matches the proper time of a classical point-like clock in the long time limit, discrepancies arise when the clock's light-crossing time is comparable to the interaction time. Moreover, even though $\mathcal{T}(C)$ matches the classical proper time of a clock in the long-time limit, we note that the proper time density $\ell_{\Omega}(\mf x)$ only reduces to the classical value in the pointlike limit. Indeed, for a spherically symmetric inertial clock, we have $\ell_\Omega(\mf x) = \text{sinc}(\Omega |\bm x|)/\tilde{F}(\Omega)$, where $\tilde{F}$ is the Fourier transform of $F$. In the pointlike limit, $\tilde{F}(\Omega)\to 1$, leading to $\ell_\Omega(\mf x) = 1$ at the trajectory $\bm x = 0$. In Appendix~\ref{sec:recover_proper}, we also show the difference between the times tracked by identical clocks following perturbed trajectories.

Finally, any operational definition of time based on a finite-sized quantum system must account for the fact that the system inevitably samples the universal vacuum fluctuations of the field. These fluctuations are not optional background contributions: in any spacetime, the short-distance singularity structure of physically admissible two-point functions is fixed by the local spacetime geometry through the Hadamard condition~\cite{fullingHadamard, bibleHadamard, fewsterNecessityHadamard}. For non-inertial trajectories, the same correlations are sampled in a trajectory-dependent way, giving rise to phenomena such as the Unruh effect~\cite{Unruh1976, Unruh-Wald, matsasUnruh, LoukoUnruh}. The clock model introduced here naturally incorporates these effects.

This is also why a clock must have a finite size. A pointlike clock would probe arbitrarily short-distance correlations of the field, leading to UV divergences in the leading-order de-excitation probability~\cite{Schlicht}. The proper spatial profile $F(\boldsymbol{\xi})$ therefore acts as a physical resolution scale, making the clock response finite while retaining the universal vacuum and non-inertial effects that any realistic quantum clock must experience.

\secl{Twin paradox in QFT} To illustrate the physical features of the clock model proposed here, we now consider a twin-paradox scenario. 
The twin paradox is usually formulated in terms of two twins, Alice and Bob, who undergo different trajectories connecting two events $\mf x_i$ and $\mf x_f$, where Alice remains inertial throughout her motion and Bob experiences acceleration. When they reunite and compare clocks, they conclude that Alice has aged more than Bob. The description of the standard setup assumes both Alice and Bob carry an ideal pointlike clock, which tracks the proper time along their trajectories and can be used to compare how much they have aged. 

If, instead, one wishes to consider more realistic finite-sized clocks, one has to be careful about how to compare the twins' motion at the initial and final events. Indeed, observers moving relative to one another have different rest spaces, and even clocks with the same shape would start their trajectory at different times within their spatial extension. Thus, when using extended clocks, it is crucial that both twins start and end their motion at rest with respect to each other. We then consider Alice undergoing the trajectory $\mf z_{\tc{a}}(t) = (t, 0, 0, 0)$ parameterized by her (inertial) proper time $t$ and Bob following the trajectory $\mf z_{\tc{b}}(\tau) = (t(\tau), x(\tau), 0, 0)$ parameterized by the (non-inertial) proper time $\tau$. His trajectory starts inertial and constantly accelerates with proper acceleration $a$ in the $x$ direction for $\tau\in(0,T/4)$, followed by a portion with constant acceleration $-a$ for $ \tau \in (T/4, 3T/4)$, and another trajectory with constant acceleration $a$ for the final part, $\tau\in (3 T/4, T)$ ending at rest with respect to Alice and meeting her again at $\tau = T$ (see Appendix~\ref{sec:calculations} for expressions). With these choices, the trajectories are initially comoving, starting their motion at the event $\mf x_i = (0,\boldsymbol{0})$ in inertial coordinates $(t,\boldsymbol{x})$. The twins meet again at the event $\mf x_f = (T_\tc{a},0)$ with
\begin{equation}
    T_{\tc{a}} = \frac{4}{a}\sinh \left(\frac{aT}{4} \right),
\end{equation}
where the trajectories are comoving once again. When they meet, a classical ideal clock carried by Alice would indicate an elapsed proper time of $T_\tc{a}$~\footnote{Notice that $T_\tc{a}$ ends up depending on $a$ and $T$, as it corresponds to the proper time at which Bob meets Alice, and Bob's trajectory is determined by these parameters. Overall, $T_\tc{a}$ and $T_\tc{b}$ are related by $T_\tc{b} = \frac{4}{a}\sinh^{-1}(aT_\tc{a}/4)$.} between events  $\mf x_i$ and $\mf x_f$, whereas the proper time measured by Bob between these events would be $T_\tc{b} = T$.

\begin{figure}[!ht]
    \centering
    \includegraphics[width=8.6cm]{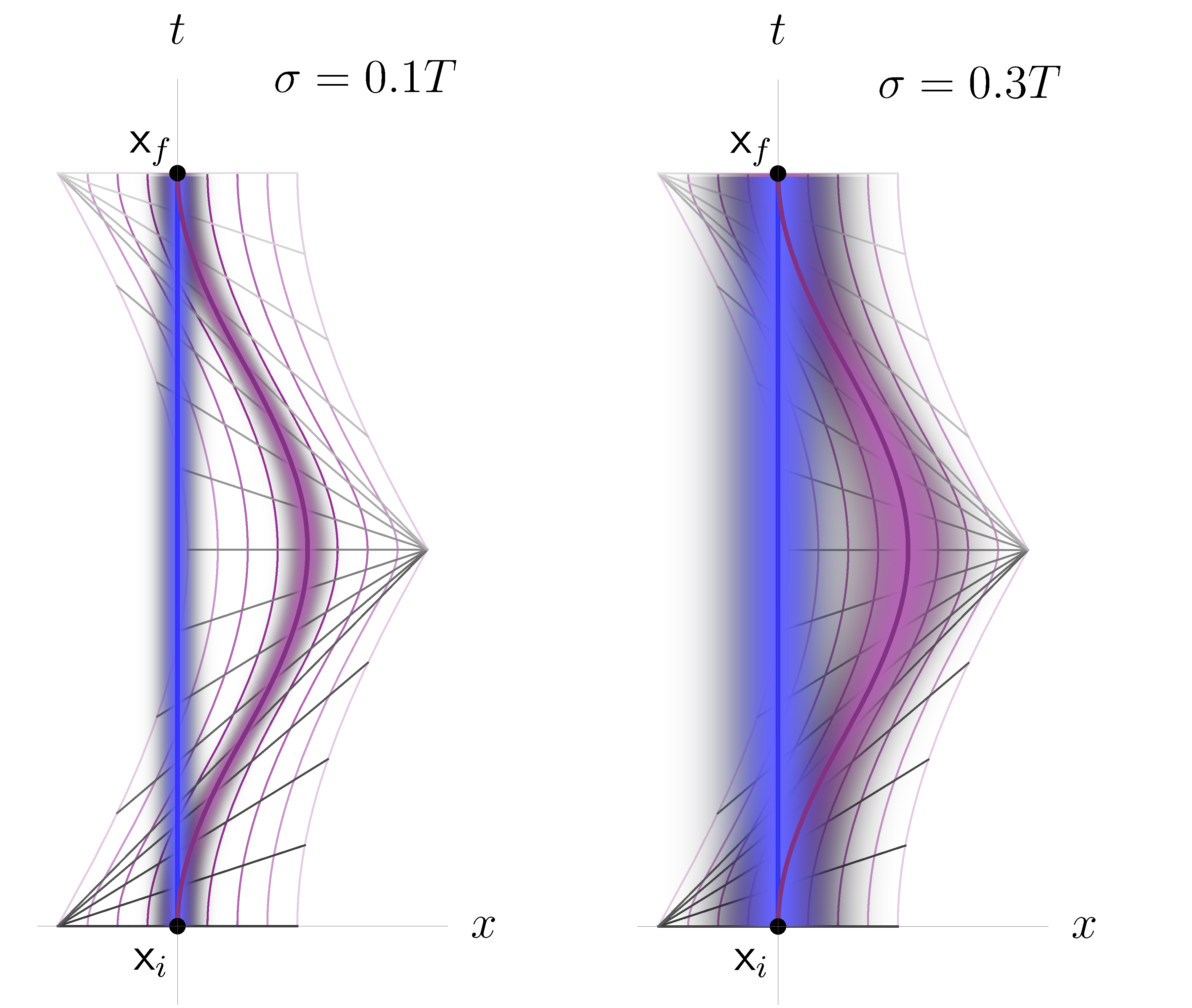}
    \caption{Density plot of the regions of interaction of Alice (blue) and Bob (purple) for $\sigma = 0.1 T$ and $\sigma = 0.3 T$. The rest surfaces of constant $\tau$ are shown in gray, and the curves of constant $\boldsymbol{\xi}$ are shown in purple.}
    \label{fig:FNC}
\end{figure}

To study the different passage of time for microscopic observers in QFT, we consider two twins, Alice and Bob, undergoing the trajectories $\mf z_\tc{a}(t)$ and $\mf z_\tc{b}(\tau)$ carrying identical clocks, as described in the previous section. The clocks are then finite-sized quantum systems with the same energy gap $\Omega$. {For mathematical convenience, we model their spatial profile as Gaussians of the form}
\begin{equation}
    F(\boldsymbol{u}) = \frac{e^{-||\boldsymbol{u}||^2/2 \sigma^2}}{(\sqrt{2 \pi} \sigma)^3},
\label{eq:gaussian_spatial_profile}
\end{equation}
where $||\boldsymbol{u}||^2 = \delta_{ij} u^i u^j$. With the choice above, $\sigma$ controls the spatial extension of the clock, also corresponding to their light-crossing time in units where $c = 1$.
Explicitly, the de-excitation probability of each clock can be written as in Eq.~\eqref{eq:P} with spacetime smearings $\Lambda_{\tc{a}}(\mf x)$ and $\Lambda_{\tc{b}}(\mf x)$ given by
\begin{align}
    \Lambda_{\tc{a}}(t, \boldsymbol{x}) &= \mathds{1}_{[0, T_{\tc{a}}]}\!(t)F(\boldsymbol{x}),\\
    \Lambda_{\tc{b}}(\tau, \boldsymbol{\xi}) &= \mathds{1}_{[0,T_{\tc{b}}]}\!(\tau)F(\boldsymbol{\xi}),
    \label{eq:lambda_BOB}
\end{align}
in their respective Fermi normal coordinates. Notice that while Alice's Fermi normal coordinates are simply $(t,\boldsymbol{x})$, Bob's coordinates $(\tau,\boldsymbol{\xi})$ are constructed from comoving coordinates for each uniformly accelerated portion of its trajectory. The coordinate lines in the $(t,x)$ plane are depicted in Fig.~\ref{fig:FNC}.

Using the clock model presented in Eq.~\eqref{eq:clock_explicit}, and noting that $\mathcal{F}_{\tc{a}} = \mathcal{F}_{\tc{b}}$ since the clocks are identical, the ratio of the twins' elapsed times between the events of departure and reunion is given by the ratio of the corresponding de-excitation probabilities:
\begin{equation}
    \frac{\mathcal{T}_{\tc{b}}(C)}{\mathcal{T}_{\tc{a}}(C)} = \frac{P_{\tc{b}}}{P_{\tc{a}}}.
\end{equation}
To understand the roles played by the QFT effects in this twin paradox scenario, this ratio should be compared with the classical ratio of proper times, namely
\begin{equation}
    \frac{T_\tc{b}}{T_\tc{a}} = \frac{aT}{4  \sinh(\tfrac{aT}{4})}.
    \label{eq:main}
\end{equation}

\begin{figure}[!ht]
    \centering
    \includegraphics[width=8.5cm]{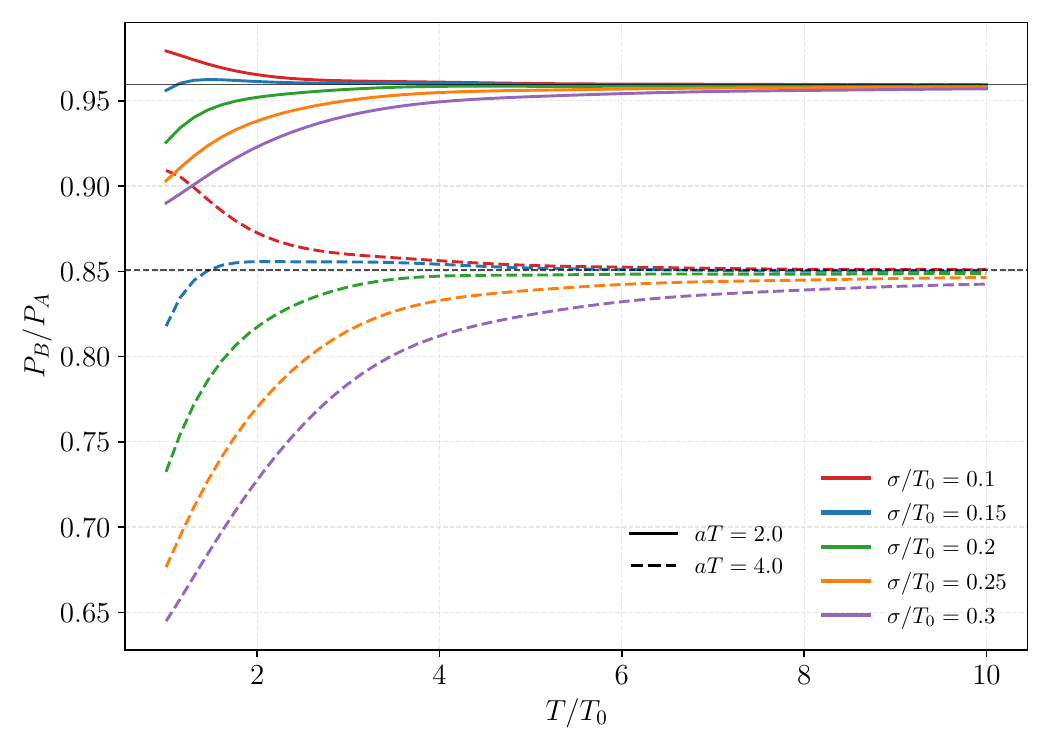}
    \caption{Ratio between Bob’s elapsed time and Alice’s elapsed time for different clock sizes (controlled by the parameter $\sigma$), with energy gap $\Omega = 2 /T_{0}$, and for trajectories satisfying $aT = 2$ and $aT = 4$.}
    \label{fig:main}
\end{figure}

\secl{Results}Using the setup above, we numerically evaluate the de-excitation probabilities \(P_{\tc{a}}\) and \(P_{\tc{b}}\) for Alice’s and Bob’s clocks, respectively. Details of the expressions used in these calculations are given in Appendix~\ref{sec:calculations}. We then compare the ratio of the clock's elapsed times, $P_{\tc{b}}/P_{\tc{a}}$, for different clock sizes and different trajectories. Observe that by fixing the value of the product $aT$, we select trajectories for which the classical proper time $T_\tc{b}/T_\tc{a}$ remains constant (as in Eq.~\eqref{eq:main}), allowing a fair comparison between different cases. Figure~\ref{fig:main} displays $P_\tc{b}/P_\tc{a}$ for multiple clock sizes with energy gap $\Omega = 2 T_0$ as a function of $T/T_0$ for a fixed timescale $T_0$ when $aT = 2$ and $aT = 4$. For the parameters used in Fig.~\ref{fig:main}, Alice's inertial clock effectively agrees with the times predicted by relativity (e.g. accuracy of $\sim 1\%$ for $\sigma=0.1T_0$).

Notice that in the long-time regime (with constant $aT$), the predictions provided by the quantum clocks agree with the classical ratio. In contrast, in the regime where the clock's light-crossing time $\sigma$ is not negligible compared to $T$, deviations appear compared to the predictions of special relativity. We observe that they (i) grow with clock size, since larger clocks sample field degrees of freedom further away from the central trajectory, and (ii) increase with the acceleration $a$, a manifestation of non-inertial QFT effects. This aligns with prior works showing that QFT features dominate at short times~\cite{quantum_clock_Jorma,quantClass}.

Moreover, the time scale at which these effects become appreciable can be compared with the precision of current atomic clocks. A single-atom atomic clock (see, e.g., Ref.~\cite{PRL_single_atom_clock2019}) has a characteristic size of order $\sigma\sim 10^{-10}\,\mathrm{m}$, and Fig.~\ref{fig:main} shows that QFT corrections become relevant when $T\sim 10\sigma/c$. For such a clock, this corresponds to a characteristic duration of order $T\sim 10^{-17}\,\mathrm{s}$. The best atomic clocks currently available can measure times of order $10^{-18}\,\mathrm{s}$~\cite{Oelker2019,PRL_single_atom_clock2019}. {Notice that this comparison is only an order-of-magnitude benchmark for the time scale at which the model predicts a visible finite-time effect; it does not imply that present atomic clocks can implement the proposed twin-paradox experiment.} Observing the effect would nonetheless require the clocks to be accelerated and decelerated through a velocity change of at least $0.45c$ on attosecond time scales, which would significantly disrupt their periodic behaviour.

In comparing our clock model with standard atomic clocks, it is also important to recall that single-atom atomic clocks usually employ a trapped ion (or neutral atom), confined by oscillatory electromagnetic fields and laser-cooled~\cite{atomic_clocks_review}. Beyond natural challenges in subjecting this setup to relativistic accelerations, the preparation of atomic clocks requires smooth Rabi pulses (to measure the current quantum state of the atom) that last typically hundreds of milliseconds~\cite{PRL_single_atom_clock2019}. For these time scales, short-time corrections are strongly suppressed, and the clock exhibits essentially classical relativistic behavior.

Overall, in experiments where atomic clocks are used to probe observer-dependent time dilation~\cite{reinhardt2007test, botermann2014test}, the acceleration (and gravitational fields) involved are much smaller than the scales where the effects discussed here are relevant. Indeed, the values used in Fig.~\ref{fig:main} would correspond to accelerations of the order of $a \sim 10^{25}\,\text{m/s}^2$ when utilizing atomic clocks to measure times of the order of $T \sim 10^{-17}\,\text{s}$, which are far beyond current experimental capability. Although the Unruh effect would be visible for these accelerations, the short duration of the interaction prevents any significant thermalization effects.



\secl{Conclusions}\label{sec: conclusion}
Time, as a physical observable, should be defined operationally in terms of physical systems, rather than introduced as a parameter of a classical theory. Although atomic clocks represent the state of the art in precision timekeeping, they rely on sustained periodic processes and are therefore not ideally suited to probe observer-dependent notions of time in regimes where significant relativistic effects occur within only a few clock cycles. Moreover, at sufficiently short time scales, physical clocks should be described within QFT, incorporating universal local vacuum fluctuations.

In this work, we introduced a finite-size clock model that incorporates these QFT effects by tracking time through the decay probability of a localized quantum system. This construction defines an effective local notion of proper time, which depends not only on the spacetime trajectory of the clock, but also on its microscopic properties, such as its spatial profile and energy gap. In the long-time regime, we showed that this effective notion of time reduces to the usual proper time of special relativity. However, in regimes where the interaction time is only a few orders of magnitude larger than the clock's proper size, the model predicts corrections arising from the way the clock samples vacuum fluctuations along its trajectory.

To further investigate these effects, we considered a twin-paradox scenario. In the long-time limit, the clocks reproduce the standard relativistic time-dilation effect. At shorter time scales, however, finite-size and QFT-induced corrections become relevant. They leave the geometric proper times unchanged, but modify the finite-time ratio registered by the clocks. This shows that, at microscopic scales, the time tracked by a physical clock is not determined solely by its trajectory. Instead, it also depends on the internal structure and dynamics of the system used to measure time. In regimes where universal QFT effects are relevant~\cite{achim2,geometry}, time becomes an operational quantity that is sensitive both to the observer's motion and to the microscopic details of the clock itself.

\acknowledgements

 MHZ thanks Profs. Achim Kempf and Eduardo Martín-Martínez for supervision, and Prof. Achim Kempf for his funding through the Dieter Schwarz grant. TRP is thankful for financial support from the Olle Engkvist Foundation (no.225-0062). Research at Perimeter Institute is supported in part by the Government of Canada through the Department of Innovation, Science and Industry Canada and by the Province of Ontario through the Ministry of Colleges and Universities. Perimeter Institute and the University of Waterloo are situated on the Haldimand Tract, land that was promised to the Haudenosaunee of the Six Nations of the Grand River, and is within the territory of the Neutral, Anishinaabe, and Haudenosaunee people. Nordita is partially supported by Nordforsk.

\newpage

\onecolumngrid

\appendix

\section{Explicit evaluation of the de-excitation probabilities for Alice and Bob}
\label{sec:calculations}

In this section, we present the expressions for the decay probability of Alice's and Bob's clocks that were used to obtain the results expressed in Figure 2 of the main text. We recall that the clock's decay probability is given by
\begin{equation}
    P = \lambda^2\int \dd V \dd V' \Lambda(\mf x) \Lambda(\mf x')e^{\ii \Omega(\tau-\tau')}W(\mf x, \mf x'),
    \label{eq:prob_decay_general}
\end{equation}
where $\tau$ is the time Fermi Normal coordinate \cite{poisson} associated with the center of mass's trajectory. The Minkowski vacuum function in $3 + 1$ dimensions for a scalar, massless field admits the following plane wave expansion,
\begin{equation}
    W(\mf x, \mf x')=\frac{1}{(2 \pi)^3} \int \frac{\dd^3 \boldsymbol{k}}{2|\boldsymbol{k}|} e^{\ii \mf k \cdot(\mf x-\mf x^{\prime})},
    \label{eq:Wightman_momentum}
\end{equation}
where $\mf k = (|\boldsymbol{k}|, \boldsymbol{k})$ and $ \mf k\cdot \mf x = \eta_{\mu \nu}k^{\mu} x^{\nu}$, with $\eta_{\mu \nu} = \text{diag}(-1, 1, 1, 1)$ the Minkowski metric. Defining the spacetime Fourier transform of the smearing as
\begin{equation}
    \tilde{\Lambda}(|\boldsymbol{k}|, \boldsymbol{k}):=\int \dd ^4  \mf x \  \Lambda(\mf x) e^{\ii \mf k\cdot \mf x},
\end{equation}
one can show that the de-excitation probability to leading order in the coupling strength $\lambda$ can be written as
\begin{equation}
    P = \frac{\lambda^2}{(2 \pi)^3}\int\frac{\dd ^3 \boldsymbol{k}}{2 |\boldsymbol{k}|}\  |\tilde{\Lambda}(|\boldsymbol{k}| + \Omega, \boldsymbol{k})|^2.
    \label{eq:prob_decay}
\end{equation}
Let us now adapt Eq.~\eqref{eq:prob_decay} to the twin‑paradox setup described in this paper. Working in inertial coordinates $\mf x=(t,\bm x)$, Alice’s clock follows the trajectory 
$ \mf {z}_{\tc{a}}(t) = (t,0)$, whereas Bob’s clock follows $\mf z_{\tc{b}}(\tau) = (t(\tau), x(\tau))$, with the parametrization in function of the proper time $\tau$ defined piecewise as follows:
\begin{align}
t(\tau) &=
\begin{cases}
\dfrac{1}{a}\sinh(a\tau),
& \tau \in (0, T/4), \\[0.7em]
\dfrac{1}{a}\sinh\!\left(a(\tau - T/2)\right)
+ \dfrac{2}{a}\sinh\!\left(\dfrac{aT}{4}\right),
& \tau \in (T/4, 3T/4), \\[0.7em]
\dfrac{1}{a}\sinh\!\left(a(\tau - T)\right)
+ \dfrac{4}{a}\sinh\!\left(\dfrac{aT}{4}\right),
& \tau \in (3T/4, T).
\end{cases}
\end{align}
\begin{align}
x(\tau) &=
\begin{cases}
\dfrac{1}{a}\cosh(a\tau) - \dfrac{1}{a},
& \tau \in (0, T/4), \\[0.7em]
-\dfrac{1}{a}\cosh\!\left(a(\tau - T/2)\right)
+ \dfrac{2}{a}\cosh\!\left(\dfrac{aT}{4}\right) - \dfrac{1}{a},
& \tau \in (T/4, 3T/4), \\[0.7em]
\dfrac{1}{a}\cosh\!\left(a(\tau - T)\right) - \dfrac{1}{a},
& \tau \in (3T/4, T).
\end{cases}
\end{align}
From the expressions above, we can also write the change from Alice's inertial coordinates $(t,\bm x) = (t,x,y,z)$ to Bob's Fermi normal coordinates $(\tau,\bm \xi) = (\tau, X, y ,z)$, namely
\begin{align}
t(\tau,X) &=
\begin{cases}
\left(X + \dfrac{1}{a}\right)\sinh(a\tau),
& \tau \in (0, T/4), \\[0.7em]
\left(-X + \dfrac{1}{a} \right)\sinh\!\left(a(\tau - T/2)\right)
+ \dfrac{2}{a}\sinh\!\left(\dfrac{aT}{4}\right),
& \tau \in (T/4, 3T/4), \\[0.7em]
\left(X + \dfrac{1}{a}\right)\sinh\!\left(a(\tau - T)\right)
+ \dfrac{4}{a}\sinh\!\left(\dfrac{aT}{4}\right),
& \tau \in (3T/4, T).
\end{cases}
\label{eq:Bob_t_of_tau}
\end{align}
\begin{align}
x(\tau,X) &=
\begin{cases}
\left(X + \dfrac{1}{a}\right)\cosh(a\tau) - \dfrac{1}{a},
& \tau \in (0, T/4), \\[0.7em]
\left(X - \dfrac{1}{a} \right)\cosh\!\left(a(\tau - T/2)\right)
+ \dfrac{2}{a}\cosh\!\left(\dfrac{aT}{4}\right) - \dfrac{1}{a},
& \tau \in (T/4, 3T/4), \\[0.7em]
\left(X + \dfrac{1}{a}\right)\cosh\!\left(a(\tau - T)\right) - \dfrac{1}{a},
& \tau \in (3T/4, T).
\end{cases}
\label{eq:Bob_x_of_tau}
\end{align}
Notice that the $y$ and $z$ coordinates remain unchanged because Bob's motion is confined to the $(t,x)$ plane. The classical proper time associated with Alice's trajectory between the events D (departure) and R (return) is then  
\begin{equation}
    T_{\tc{a}} = t(T) = \frac{4}{a}\sinh \left(\frac{aT}{4} \right).
\end{equation}
Now, recall that Alice's clock is modelled using the spacetime smearing
\begin{equation}
    \Lambda_{\tc{a}}(t, \bm x) = \mathds{1}_{[0, T_{\tc{a}}]}(t)F(\bm x),
    \label{eq:Lambda_A_appendix}
\end{equation}
with a Gaussian spatial profile,
\begin{equation}
    F(\bm x) = \frac{e^{-(x^2 + y^2 + z^2)/2 \sigma^2}}{(\sqrt{2 \pi} \sigma)^3},
\label{eq:gaussian_spatial_profile_appendix}
\end{equation}
 The evaluation of Alice's decay probability, denoted by $P_{\tc{a}}$, is then a straightforward application of Eq.~\eqref{eq:prob_decay} using the smearing defined by Eq.~\eqref{eq:Lambda_A_appendix}. After evaluating the Fourier transform of the smearing $\Lambda_{\tc{a}}$, we write the integral in momentum space using spherical coordinates. The integral over the angular variables can be evaluated, leaving just one integral that needs to be evaluated numerically. The result is
\begin{align}
P_{\tc{a}} = \lambda^2\int_{0}^{\infty}{\dd \omega
\frac{e^{-\sigma^2 \omega^2}\,\omega\,\sin^2\!\bigl(\tfrac{2(\omega - \Omega)\sinh(aT/4)}{a}\bigr)}
{\pi^2\,(\omega - \Omega)^2}
} &= \frac{4 \lambda^2\sinh^2(aT/4)}{a^2}\int_{0}^{\infty}{\dd \omega
\frac{e^{-\sigma^2 \omega^2}\,\omega\,\sin^2\!\bigl(\tfrac{2(\omega - \Omega)\sinh(aT/4)}{a}\bigr)}
{4\pi^2\,(\omega - \Omega)^2\sinh^2(aT/4)/a^2}
}\\
& = \frac{4 \lambda^2\sinh^2(aT/4)}{\pi^2 a^2}\int_{0}^{\infty}{\dd \omega
e^{-\sigma^2 \omega^2}\,\omega\,\text{sinc}^2\!\bigl(\tfrac{2(\omega - \Omega)\sinh(aT/4)}{a}\bigr)
}.
\label{eq:P_A_numerics}
\end{align}
As for the evaluation of Bob's decay probability, denoted $P_{\tc{b}}$, we use the smearing
\begin{equation}
    \Lambda_{\tc{b}}(\tau, \bm \xi) = \mathds{1}_{[0, T_{\tc{b}}]}(\tau)F(\bm \xi),
\end{equation}
where $T_{\tc{b}} = T$ is the proper time associated with the trajectory $\mf z_{\tc{b}}$ between the events D and R, and $(\tau, \bm \xi) = (\tau, X, y, z)$ are the Fermi normal coordinates associated with the timelike trajectory $\mf z_{\tc{b}}$. Thus, defining $\widetilde{W}(\tau,X,y,z; \tau', X', y', z') = W(t(\tau,X),x(\tau,X),y,z; t'(\tau',X'),x(\tau',X'),y',z')$, we can write
\begin{equation}
    P_{\tc{b}} = \lambda^2 \int_{0}^{T}\dd \tau \int_{0}^{T}\dd \tau' \int \dd X \dd y \dd z \dd X' \dd y' \dd z' e^{\ii \Omega (\tau - \tau')}F(X, y, z)F(X', y', z')\widetilde{W}(\tau,X,y,z; \tau', X', y', z') ,
    \label{eq:P_b_cumbersome}
\end{equation}
where $t(\tau, X)$ and $x(\tau, X)$ are given by Eqs.~\eqref{eq:Bob_t_of_tau} and \eqref{eq:Bob_x_of_tau}, depending on the value of the parameter $\tau$, and the spatial profile $F$ is given by Eq.~\eqref{eq:gaussian_spatial_profile_appendix}. To express Eq.~\eqref{eq:P_b_cumbersome} in a form suitable for
numerical evaluation, we substitute the corresponding expressions for
$F$, $t(\tau,X)$, and $x(\tau,X)$ in each of the three segments of
Bob's trajectory and partition the $(\tau,\tau')$ integration domain
accordingly. This procedure yields nine contributions. Hermiticity of
the Wightman function pairs the six off-diagonal contributions into
three complex-conjugate pairs, allowing Bob's de-excitation probability to be written
as
\begin{equation}
    P_{\tc{b}} = 
  P_{11} + P_{22} + P_{33} + 2(\text{Re}(P_{12}) + \text{Re}(P_{13}) + \text{Re}(P_{23})),
\end{equation}
To obtain an explicit expression for each term, we substitute the
plane-wave expansion of the Wightman function,
Eq.~\eqref{eq:Wightman_momentum}, and introduce spherical coordinates
in momentum space. The resulting terms can be written as
\begin{equation}
    P_{11} = \lambda^2\int_{0}^{T/4}\dd \tau \int_{0}^{T/4}\dd \tau' \int_{0}^{\pi}\dd \theta \int_{0}^{\infty} \dd \omega f_{11}(\tau, \tau', \theta, \omega),
\end{equation}
\begin{equation}
    P_{22} = \lambda^2\int_{T/4}^{3T/4}\dd \tau \int_{T/4}^{3T/4}\dd \tau' \int_{0}^{\pi}\dd \theta \int_{0}^{\infty} \dd \omega f_{22}(\tau, \tau', \theta, \omega),
\end{equation}
\begin{equation}
    P_{33} = \lambda^2\int_{3T/4}^{T}\dd \tau \int_{3T/4}^{T}\dd \tau' \int_{0}^{\pi}\dd \theta \int_{0}^{\infty} \dd \omega f_{33}(\tau, \tau', \theta, \omega),
\end{equation}
\begin{equation}
    P_{12} =\lambda^2 \int_{0}^{T/4}\dd \tau \int_{T/4}^{3T/4}\dd \tau' \int_{0}^{\pi}\dd \theta \int_{0}^{\infty} \dd \omega f_{12}(\tau, \tau', \theta, \omega),
\end{equation}
\begin{equation}
    P_{13} = \lambda^2\int_{0}^{T/4}\dd \tau \int_{3T/4}^{T}\dd \tau' \int_{0}^{\pi}\dd \theta \int_{0}^{\infty} \dd \omega f_{13}(\tau, \tau', \theta, \omega),
\end{equation}
\begin{equation}
    P_{23} = \lambda^2\int_{T/4}^{3T/4}\dd \tau \int_{3T/4}^{T}\dd \tau' \int_{0}^{\pi}\dd \theta \int_{0}^{\infty} \dd \omega f_{23}(\tau, \tau', \theta, \omega).
\end{equation}
In the integrals above, the functions $f_{ij}(\tau, \tau', \theta, \omega)$ can be explicitly written as
\begin{equation}
\begin{aligned}
f_{11}(\tau, \tau', \theta, \omega) 
&= \frac{1}{8 \pi^{2}} 
\exp\Bigg[
\frac{1}{8} \Big(
    -\sigma^{2} \omega^{2}
    + 8 i (\tau - \tau') \, \Omega
    + 3 \sigma^{2} \omega^{2} \cos(2\theta) \\
&\quad - \frac{\omega}{a} \Big(
    -8 i \cos\theta \, \cosh(a \tau)
    + 4 a \sigma^{2} \omega \cos^{2}\theta \, \cosh^{2}(a \tau)
    + 8 i \big( \cos\theta \, \cosh(a \tau') + \sinh(a \tau) - \sinh(a \tau') \big) \\
&\qquad\quad
    + a \sigma^{2} \omega \big(
        2 \cosh(2 a \tau)
        + (3 + \cos(2\theta)) \cosh(2 a \tau')
        - 4 \cos\theta \big( \sinh(2 a \tau) + \sinh(2 a \tau') \big)
      \big)
  \Big)
\Big)
\Bigg] \, \omega \sin\theta,
\end{aligned}
\end{equation}
\begin{equation}
\begin{aligned}
f_{22}(\tau, \tau', \theta, \omega) 
&= \frac{1}{8 \pi^{2}} 
\exp\Bigg[
    i (\tau - \tau') \, \Omega
    + \frac{\omega}{8 a} \Big(
        -8 i \cos\theta \, \cosh\!\left(\tfrac{a}{2} (T - 2\tau)\right) \\
&\quad\quad
        + 8 i \Big( 
            \cos\theta \, \cosh\!\left(\tfrac{a}{2} (T - 2\tau')\right)
            + \sinh\!\left(\tfrac{a}{2} (T - 2\tau)\right)
            - \sinh\!\left(\tfrac{a}{2} (T - 2\tau')\right)
        \Big) \\
&\quad\quad
        - a \sigma^{2} \omega \Big(
            (3 + \cos 2\theta) \, \cosh\!\left(a (T - 2\tau)\right)
            + (3 + \cos 2\theta) \, \cosh\!\left(a (T - 2\tau')\right) \\
&\qquad\quad\quad
            + 4 \sin^{2}\theta
            - 4 \cos\theta \big(
                \sinh\!\left(a (T - 2\tau)\right)
                + \sinh\!\left(a (T - 2\tau')\right)
            \big)
        \Big)
    \Big)
\Bigg] \, \omega \sin\theta
\end{aligned}
\end{equation}
\begin{equation}
\begin{aligned}
f_{33}(\tau, \tau', \theta, \omega)
&= \frac{1}{8 \pi^{2}}
\exp\Bigg[
    i (\tau - \tau') \, \Omega
    - \frac{\sigma^{2} \omega^{2}}{8} \Big(
        (3 + \cos 2\theta) \cosh\!\left( 2a (T - \tau) \right)
        + (3 + \cos 2\theta) \cosh\!\left( 2a (-T + \tau') \right) \\
&\quad\quad
        + 4 \big( \sin^{2}\theta
                  + \cos\theta \big(
                      \sinh\!\left( 2a (T - \tau) \right)
                      + \sinh\!\left( 2a (T - \tau') \right)
                  \big)
          \big)
    \Big) \\
&\quad\quad
    + \frac{i \omega}{a} \Big(
        \cos\theta \big(
            \cosh\!\left( a (-T + \tau) \right)
            - \cosh\!\left( a (-T + \tau') \right)
        \big)
        + \sinh\!\left( a (T - \tau) \right)
        + \sinh\!\left( a (-T + \tau') \right)
    \Big)
\Bigg] \, \omega \sin\theta
\end{aligned}
\end{equation}
\begin{equation}
\begin{aligned}
f_{12}(\tau, \tau', \theta, \omega)
&= \frac{1}{8 \pi^{2}}
\exp\Bigg[
\frac{1}{8} \Big(
    - 2\sigma^{2} \omega^{2}
    + 8 i (\tau - \tau') \, \Omega
    + 2\sigma^{2} \omega^{2} \cos 2\theta \\
&\quad\quad
    + \frac{\omega}{a} \Big(
        - a\sigma^{2} \omega (3 + \cos 2\theta) \big[
            \cosh(2 a \tau) + \cosh(a (T - 2\tau'))
        \big] \\
&\qquad\quad
        + 8 i \big[
            2\sinh\!\left(\frac{a T}{4}\right)
            - \sinh(a \tau)
            - \sinh\!\left(\frac{a}{2} (T - 2\tau')\right)
        \big] \\
&\qquad\quad
        + 4\cos\theta \big[
            -4 i \cosh\!\left(\frac{a T}{4}\right)
            + 2 i \cosh(a \tau)
            + 2 i \cosh\!\left(\frac{a}{2} (T - 2\tau')\right) \\
&\hspace{4.8cm}
            + a\sigma^{2} \omega \big(
                \sinh(2 a \tau)
                + \sinh(a (T - 2\tau'))
            \big)
        \big]
    \Big)
\Big)
\Bigg] \, \omega \sin\theta
\end{aligned}
\end{equation}
\begin{equation}
\begin{aligned}
f_{13}(\tau, \tau', \theta, \omega)
&= \frac{1}{8 \pi^{2}}
\exp\Bigg[
\frac{1}{8} \Big(
    - \sigma^{2} \omega^{2}
    + 8 i (\tau - \tau') \, \Omega
    + 3\sigma^{2} \omega^{2} \cos 2\theta \\
&\quad\quad
    - \frac{\omega}{a} \Big(
        -8 i \cos\theta \, \cosh(a \tau)
        + 4 a\sigma^{2} \omega \cos^{2}\theta \, \cosh^{2}(a \tau) \\
&\qquad\quad
        + 8 i \big[
            \cos\theta \, \cosh(a (-T + \tau'))
            - 4\sinh\!\left(\frac{a T}{4}\right)
            + \sinh(a \tau)
            + \sinh(a (T - \tau'))
        \big] \\
&\qquad\quad
        + a\sigma^{2} \omega \big[
            2\cosh(2 a \tau)
            + (3 + \cos 2\theta) \cosh\!\left(2 a (-T + \tau')\right) \\
&\hspace{5.2cm}
            - 4\cos\theta \big(
                \sinh(2 a \tau)
                + \sinh\!\left(2 a (-T + \tau')\right)
            \big)
        \big]
    \Big)
\Big)
\Bigg] \, \omega \sin\theta
\end{aligned}
\end{equation}
\begin{equation}
\begin{aligned}
f_{23}(\tau, \tau', \theta, \omega)
&= \frac{1}{8 \pi^{2}}
\exp\Bigg[
\frac{1}{8} \Big(
    - 2\sigma^{2} \omega^{2}
    + 8 i (\tau - \tau') \, \Omega
    + 2\sigma^{2} \omega^{2} \cos 2\theta \\
&\quad\quad
    + \frac{\omega}{a} \Big(
        - a\sigma^{2} \omega (3 + \cos 2\theta) \big[
            \cosh\!\left(a (T - 2\tau)\right)
            + \cosh\!\left(2 a (-T + \tau')\right)
        \big] \\
&\qquad\quad
        + 8 i \big[
            2\sinh\!\left(\frac{a T}{4}\right)
            + \sinh\!\left(\frac{a}{2} (T - 2\tau)\right)
            + \sinh\!\left(a (-T + \tau')\right)
        \big] \\
&\qquad\quad
        + 4\cos\theta \big[
            4 i \cosh\!\left(\frac{a T}{4}\right)
            - 2 i \cosh\!\left(\frac{a}{2} (T - 2\tau)\right)
            - 2 i \cosh\!\left(a (-T + \tau')\right) \\
&\hspace{5.3cm}
            + a\sigma^{2} \omega \big(
                \sinh\!\left(a (T - 2\tau)\right)
                + \sinh\!\left(2 a (-T + \tau')\right)
            \big)
        \big]
    \Big)
\Big)
\Bigg] \, \omega \sin\theta
\end{aligned}
\end{equation}

\section{Recovering ideal clocks in the long-time regime}
\label{sec:inertial_clocks}

In this section, we analyze the behaviour of our decay-based clock
$C$ along both inertial and non-inertial trajectories. For a
spherically symmetric spatial profile,
$F(\bm x)=F(|\bm x|)$, we show that the model approaches the
predictions of an ideal clock when the clock's light-crossing time is
negligible compared with the duration being measured.

Consider first a clock following an inertial trajectory in Minkowski
spacetime and interacting with the field for a proper time $T$. For
the switching function
$\chi(t)=\mathds{1}_{[0,T]}(t)$, the de-excitation probability at
leading nonvanishing order in the coupling strength is
\begin{equation}
    P(T) = \left(\frac{\lambda T}{2\pi}\right)^2
\int_{0}^{\infty} \dd\omega\,\tilde{F}(\omega)^2\,\omega\,
\operatorname{sinc}^2\!\left(\frac{(\omega - \Omega)T}{2}\right),
\label{eq:PTsinc}
\end{equation}
where $\tilde{F}(\omega) = \tilde{F}(\omega)^*$ is the Fourier transform of $F(|\bm x|)$.  Now, by defining the variable $u = \omega T$, and then performing a translation $u \to u + \Omega T$, we can write
\begin{align}
P(T) &= \frac{\lambda^2}{(2\pi)^2}
\int_{0}^{\infty} \dd u\, \tilde{F}(u/T)^2\,u\,
\operatorname{sinc}^2\!\left(\frac{(u - \Omega T)}{2}\right), \nonumber \\
&= \frac{\lambda^2}{(2\pi)^2}
\int_{-\Omega T}^{\infty} \dd u\, \tilde{F}(u/T+\Omega)^2\,(u+ \Omega T)\,
\operatorname{sinc}^2\!\left(\frac{u}{2}\right)
\end{align}
 Then, to leading order in the coupling strength $\lambda$, the average de-excitation rate can be cast into
\begin{equation}
\frac{P(T)}{T} = \frac{
\lambda^2
}{(2\pi)^2}
\int_{-\Omega T}^{\infty} \dd u\, \tilde{F}(u/T+\Omega)^2\,(\tfrac{u}{T}+ \Omega)\,
\operatorname{sinc}^2\!\left(\frac{u}{2}\right).
\end{equation}
We now evaluate the long-interaction-time limit $T\to\infty$ at fixed
spatial width $\sigma$. Equivalently, this is the regime
$\sigma/T\to0$, in which the clock's light-crossing time is negligible
compared with the duration being measured. We obtain
\begin{align}
\mathcal{F} = \lim_{T\to \infty}\frac{P(T)}{T} &=  \lim_{T\to \infty}\frac{\lambda^2}{(2\pi)^2}
\int_{-\infty}^{\infty} \dd u\, \tilde{F}(u/T+\Omega)^2\,\left(\tfrac{u}{T}+ \Omega\right)\,
\operatorname{sinc}^2\!\left(\frac{u}{2}\right) \nonumber \\
&=  \lim_{T\to \infty}\frac{\lambda^2}{(2\pi)^2}\left(\Omega
\int_{-\infty}^{\infty} \dd u\, \tilde{F}(u/T+\Omega)^2\,
\operatorname{sinc}^2\!\left(\frac{u}{2}\right) + 
\frac{1}{T}\int_{-\infty}^{\infty} \dd u\, \tilde{F}(u/T+\Omega)^2\, u
\operatorname{sinc}^2\!\left(\frac{u}{2}\right)\right) \nonumber\\
&=  \lim_{T\to \infty}\frac{\lambda^2\Omega}{(2\pi)^2}
\int_{-\infty}^{\infty} \dd u\,\tilde{F}(u/T+\Omega)^2\,
\operatorname{sinc}^2\!\left(\frac{u}{2}\right)\nonumber\\
&=  \frac{ \lambda^2\Omega}{(2\pi)^2}
\int_{-\infty}^{\infty} \dd u\, \lim_{T\to \infty}\tilde{F}(u/T+\Omega)^2\,
\operatorname{sinc}^2\!\left(\frac{u}{2}\right)\nonumber\\
&=\frac{\lambda^2\Omega \tilde{F}(\Omega)^2}{(2\pi)^2}\int_{-\infty}^\infty \dd u \,\text{sinc}^2(u/2) \nonumber\\
&= \frac{\lambda^2\Omega\tilde{F}(\Omega)^2}{2\pi},
\label{eq:ideal_clock}
\end{align}
where we assume that $\tilde{F}(u)$ is bounded ($|\tilde{F}(u)|\leq \alpha$), so that we can commute the limit with the integral using the dominated convergence theorem due to the fact that $\tilde{F}(u/T+\Omega)^2\text{sinc}^2(u/2)\leq \alpha^2 \, \text{sinc}^2(u/2)$. 

\begin{figure}[!ht]
    \centering
    \includegraphics[width=0.48\textwidth]{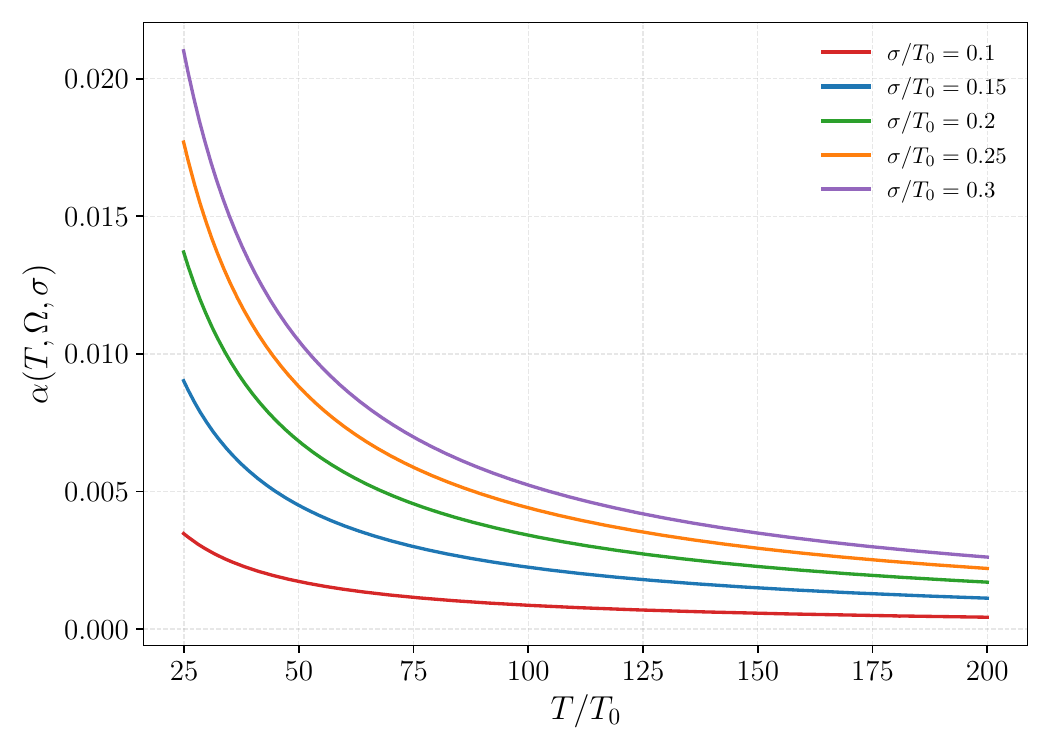}
    \caption{Relative deviation $\alpha(T, \Omega, \sigma)$ between our model and an ideal clock in an inertial scenario, for different clock sizes (as controlled by the parameter $\sigma$) and with energy gap $\Omega =2 /T_{0}$.}
    \label{fig:inertial}
\end{figure}

We now pick $F(|\bm x|)$ as the normalized Gaussian chosen in the main text, so that $\tilde{F}(\Omega) = e^{- \frac{\sigma^2\Omega^2}{2}}$, yielding the following de-excitation rate
\begin{equation}
    \mathcal{F} = \frac{\lambda^2 \Omega e^{- \sigma^2 \Omega^2}}{2\pi}.
\end{equation}
Notice that the rate above remains finite as $\sigma\to 0$, even
though the de-excitation probability diverges in this limit; see
Eq.~\eqref{eq:PTsinc}. Thus, the limits $T\to\infty$ and
$\sigma\to0$ do not commute. Equation~\eqref{eq:ideal_clock} shows
that, in the limit $T\to\infty$, the QFT-based clock $C$ behaves as an
ideal clock, with the de-excitation rate $\mathcal{F}$ providing the
proportionality constant between the decay probability and the
measured time. For a finite interaction time $T$, the time registered
by the clock is defined according to Eq.~\eqref{eq:clock_explicit} of the main text. In
this case, we denote $\mathcal{T}(C)$ by $\Delta\tau(T)$, thus writing
\begin{equation}
    \Delta \tau(T)= \frac{P(T)}{\mathcal{F}}.
\end{equation}
Equivalently, the relative deviation from the ideal clock case is measured by the quantity
\begin{equation}
    \alpha(T, \Omega, \sigma) \equiv \frac{|P(T)/T - \mathcal{F}|}{\mathcal{F}}.
    \label{eq:clock_def_appendix_2}
\end{equation}
In Figure \ref{fig:inertial}, we display the values of the relative deviation $\alpha(T, \Omega, \sigma)$ for different clocks' sizes with energy gap $\Omega = 2/T_{0}$, for some fixed timescale $T_{0}$. We observe that, when the clock's size is about $10^{3}$ smaller than the times being measured, the relative error compared to the ideal clock is less than $0.3 \%$. For the case used to explore the twin paradox scenario, Alice's clock is at most of size $\sigma = 0.3 T_0$ and for $T = 10T_0$, the deviation from the standard clock constant is at most $5\%$, being $1\%$ for the smallest clock size considered of $\sigma = 0.1T_0$.

\begin{figure}[!htb]
    \centering
    \includegraphics[width=\linewidth]{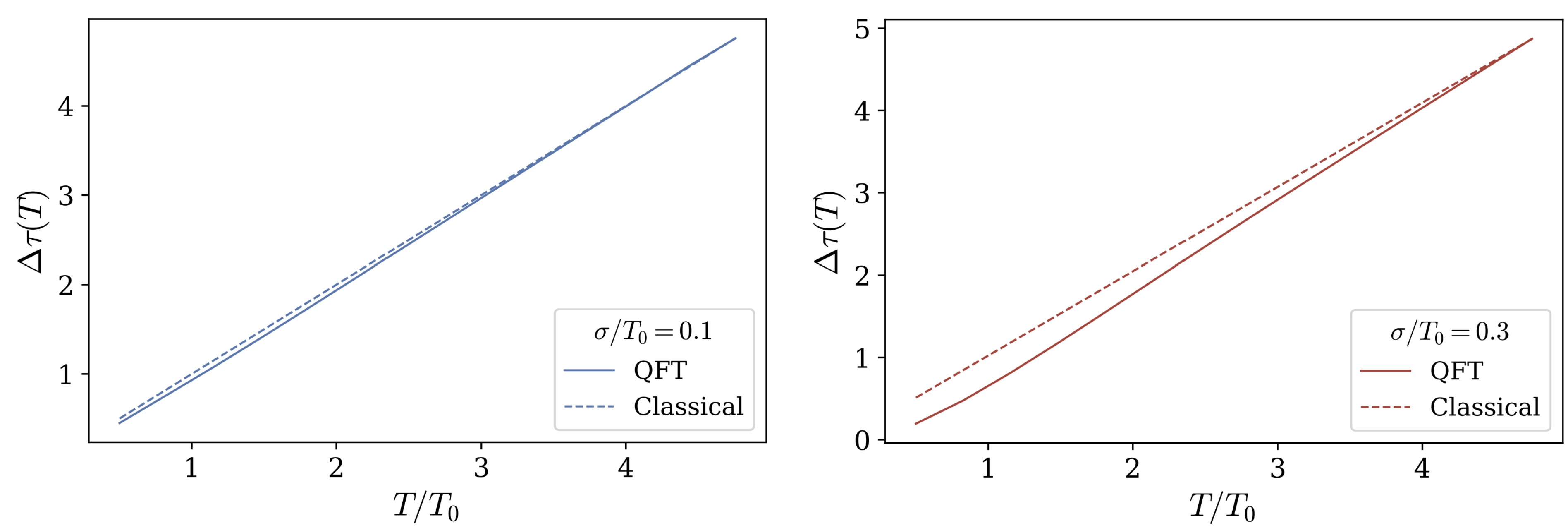}
\caption{
Finite-time behavior of the QFT-based clock model for non-inertial trajectories with $aT_{0}=2$. The clock gap is fixed by $\Omega T_{0}=2$, and we compare two proper sizes, $\sigma=0.1T_{0}$ and $\sigma=0.3T_{0}$. In both cases, as the total interaction time $T$ increases, the time registered by the clock, $\Delta\tau(T)$, approaches the ideal classical behavior $\Delta\tau(T)\simeq T$.
}
    \label{fig:nonInertial}
\end{figure}

For non-inertial clocks, the same reasoning suggests that the clock
should approach ideal-clock behaviour when its light-crossing time is
negligible compared with the duration being measured. To illustrate
this, we consider Bob's non-inertial trajectory, defined in
Eqs.~\eqref{eq:Bob_t_of_tau}--\eqref{eq:Bob_x_of_tau}. For this
trajectory, we estimate the rate $\mathcal{F}$ numerically and then
use Eq.~\eqref{eq:clock_def_appendix} to compute the registered time
$\Delta\tau(T)$. Figure~\ref{fig:nonInertial} shows
$\Delta\tau(T)$ for trajectories with proper acceleration fixed by
$aT_0=2$. We consider two clock sizes, $\sigma=0.1T_0$ and
$\sigma=0.3T_0$, and set the energy gap to $\Omega=2/T_0$.

\section{Perturbation of timelike trajectories in FNC}
\label{sec:FNC_perturbation_justified}
Here, we introduce a construction in Fermi normal coordinates (FNC)
that will be used to derive the results of the next section. Consider
a reference timelike curve $\mf z(\tau)$ in Minkowski spacetime,
parametrized by its proper time $\tau$, and let $(\tau,\xi^i)$ denote
the associated FNC. At each point $\mf z(\tau)$, let
$e^\mu{}_{a}(\tau)$ be an orthonormal tetrad:
\begin{equation}
    \eta_{\mu \nu}e^{\mu}_{\ a}e^{\nu}_{\ b} = \eta_{ab}
\end{equation}
Denoting the tangent to the curve by $u(\tau)$, we take
\begin{equation}
    e_{0}(\tau) = u(\tau), \quad e^{\mu}_{\ i}(\tau) u_{\mu}(\tau) = 0.
\end{equation}
The tetrad is transported along the reference curve according to the Fermi-Walker transport:
\begin{equation}
    \frac{d e^{\mu}_{\ i}}{d \tau} = a_{i}(\tau) u^{\mu}(\tau), \quad a_{i}(\tau) = e^{\mu}_{\ i}(\tau)a_{\mu}(\tau).
    \label{eq:FW_transport}
\end{equation}
Let $x^{\mu} = (t, x^{i})$ denote global inertial coordinates. Then, we have
\begin{equation}
    x^{\mu} = z^{\mu}(\tau) + \xi^{i}e^{\mu}_{\ i}(\tau).
    \label{eq:coordinates_inertial_F}
\end{equation}
Now, consider another reference timelike curve, $\tilde{\mf z}(\tilde{\tau})$. Let the associated FNC be denoted by $(\tilde{\tau}, \tilde{\xi}^{i})$. This new curve is supposed to be a perturbation of the first, with the same end points $\mf x_{f} = \mf z(\tau_{f})$ and $\mf x_{i} = \mf z(\tau_{i})$. By assumption,
\begin{equation}
    \tilde{\tau} = \tau + \delta \tau(\tau, \boldsymbol{\xi}),
\end{equation}
\begin{equation}
    \tilde{\xi}^{i} = \xi^{i} + \delta \xi^{i}(\tau, \boldsymbol{\xi}).
\end{equation}
Because each curve is parametrized by its own proper time, their
points cannot be compared directly. We therefore express the proper
time along the second reference curve in terms of $\tau$ as
\begin{equation}
\tilde{\tau}_{\text{curve}} = \tau + \alpha(\tau),
\end{equation}
with $\alpha(\tau)$ small enough. This way, we can define
\begin{equation}
    \delta z^{\mu}(\tau)  \equiv \tilde{z}^{\mu}(\tau + \alpha(\tau)) - z^{\mu}(\tau).
\end{equation}
Then, to leading order we can write
\begin{align}
    \tilde{z}^{\mu}(\tilde{\tau})&  = \tilde{z}^{\mu}(\tau + \alpha) + \tilde{u}^{\mu}(\tau + \alpha)(\delta \tau - \alpha) \nonumber \\ & = \delta z^{\mu}(\tau) + z^{\mu}(\tau) + \tilde{u}^{\mu}(\tau + \alpha)(\delta \tau - \alpha) \nonumber \\ & = 
    \delta z^{\mu}(\tau) + z^{\mu}(\tau) + u^{\mu}(\tau)(\delta \tau - \alpha).
\end{align}
Similarly,
\begin{equation}
    \tilde{e}^{\mu}_{\ i}(\tilde{\tau}) = e^{\mu}_{\ i}(\tau) + \delta e^{\mu}_{\ i}(\tau) + a_{i}u^{\mu}(\delta \tau - \alpha),
\end{equation}
where we used Eq.~\eqref{eq:FW_transport}. Next, using Eq.~\eqref{eq:coordinates_inertial_F}, it follows that
\begin{equation}
        z^{\mu}(\tau) + \xi^{i}e^{\mu}_{\ i}(\tau) = \tilde{z}^{\mu}(\tilde{\tau}) + \tilde{\xi}^{i} \tilde{e}^{\mu}_{ \ i}(\tilde{\tau}).
\end{equation}
Using the expansion to leading order, this simplifies as
\begin{equation}
    \delta z^{\mu} + u^{\mu}(\delta \tau - \alpha)(\xi^{i}a_{i} + 1) + \xi^{i}\delta e^{\mu}_{\ i} + \delta \xi^{i}e^{\mu}_{\ i} = 0.
    \label{eq:another_day_another_equation}
\end{equation}
Then, contracting with $u_{\mu}$, we obtain
\begin{equation}
    \delta \tau = \frac{u_{\mu}(\delta z^{\mu} + \xi^{i}\delta e^{\mu}_{ \ i}) + \alpha(\xi^{i}a_{i} + 1)}{\xi^{i}a_{i} + 1}.
\end{equation}
Finally, Substituting this into Eq.~\eqref{eq:another_day_another_equation} yields
\begin{equation}
    \delta \xi^{i} e^{\mu}_{\ i} = -\delta z^{\mu}.
\end{equation}
Hence,
\begin{equation}
    \delta \xi^{i} = - \delta z^{\mu}e_{\mu}^{ \ i}.
\end{equation}

\section{Recovering a notion of proper time density for QFT-based clocks}
\label{sec:recover_proper}
The notion of an ideal clock in general relativity can be extended to
clocks with finite spatial extent. Such a clock is specified not only
by a reference trajectory $\mf z(\tau)$, but also by a spatial profile
$F(\bm\xi)$, normalized to unity, where $(\tau,\bm\xi)$ are the Fermi
normal coordinates associated with the trajectory. The time measured
by the clock $C=(\mf z(\tau),F(\bm\xi))$ can then be written as
\begin{equation}
    \mathcal{T}(C_{\text{ideal}}) = \int_{\tau_{i}}^{\tau_{f}} \dd \tau\int \dd^{3} \boldsymbol{\xi} \, F(\boldsymbol{\xi})  \ell(\tau, \boldsymbol{\xi}),
    \label{eq:ideal_smeared_appendix}
\end{equation}
where
\begin{equation}
    \ell (\tau, \boldsymbol{\xi}) = \sqrt{-g_{\mu \nu}(\mf z_{\boldsymbol{\xi}}(\tau)) \dv{z_{\bm \xi}^{\mu}}{\tau}\dv{z_{\bm \xi}^{\nu}}{\tau}}
\end{equation}
is the proper time density within the clock's world tube, where $z_{\bm \xi}$ are the timelike trajectories that compose such a tube. Explicitly, we can write
\begin{equation}
    z_{\bm \xi}^\mu(\tau) = (\tau,\bm \xi), \quad\quad \dv{z_{\bm \xi}^\mu}{\tau} = (1,\bm 0) \quad\quad \Rightarrow \quad\quad \ell(\tau,\bm \xi) = \sqrt{-g_{00}(\tau,\bm \xi)}.
\end{equation}
Moreover, the integral in Eq.~\eqref{eq:ideal_smeared_appendix} can be rewritten as a four-dimensional integral by letting $\mf x = (\tau, \bm \xi)$. Explicitly,
\begin{equation}
    \mathcal{T}(C_{\text{ideal}}) = \int \dd^{4} \mf x \, F(\bm \xi)\chi(\tau) \ell(\mf x),
\end{equation}
with 
\begin{equation}
    \chi(\tau) = \Theta(\tau-\tau_i) - \Theta(\tau-\tau_{f}),
\end{equation}
 and
\begin{equation}
    \ell(\mf x) = \sqrt{-g_{\mu \nu}(\mf x)u^{\mu}(\mf x)u^{\nu}(\mf x)},
\end{equation}
where $u^\mu(\mf x)$ denotes the tangent vector to the timelike
trajectory $z_{\bm\xi}$ passing through $\mf x$ within the clock's
world tube.

Suppose now that the same clock follows a reference trajectory
$\tilde{\mf z}(\tau)$ obtained by perturbing $\mf z(\tau)$. We ask how
this perturbation changes the measured time
$\mathcal{T}(C_{\text{ideal}})$. Using the construction introduced in
Sec.~\ref{sec:FNC_perturbation_justified}, let $\delta u^\mu$ denote
the corresponding perturbation of the tangent field. To first order
in the perturbation, we have
\begin{equation}
    \tilde{\ell}(\mf x) = \sqrt{-g_{\mu \nu}(u^{\mu}u^{\nu} + 2 u^{\mu} \delta u^{\nu})} = \sqrt{-g_{\mu \nu}u^{\mu}u^{\nu}}\sqrt{ 1 + \frac{2 g_{\mu \nu}u^{\mu}\delta u^{\nu}}{g_{\mu \nu}u^{\mu}u^{\nu}}} = \ell(\mf x) - \frac{g_{\mu \nu}u^{\mu}\delta u^{\nu}}{\ell(\mf x)} + \mathcal{O}(\delta^2).
\end{equation}
Hence,
\begin{align}
    \mathcal{T}(\tilde{C}_{\text{ideal}}) & = \int \dd^{4}\mf x\,  F(\tilde{\bm \xi})\tilde{\chi}(\tilde{\tau}) \tilde{\ell}(\mf x)  \nonumber\\ & = \int \dd^{4}\mf x\,  F(\bm \xi + \delta \bm \xi)\tilde{\chi}(\tau + \delta \tau)\left(\ell(\mf x) -  \frac{g_{\mu \nu}u^{\mu}\delta u^{\nu}}{\ell(\mf x)}\right) \nonumber \\ & = \mathcal{T}(C_{\text{ideal}})  - \int \dd^{4} \mf x\,  F(\bm \xi)\tilde{\chi}(\tau) \frac{g_{\mu \nu}u^{\mu}\delta u^{\nu}}{\ell(\mf x)}  + \int \dd^{4} \mf x \, \delta \bm \xi \cdot \nabla F(\bm \xi) \tilde{\chi}(\tau)\ell(\mf x) + \mathcal{O}(\delta^2) \nonumber \\ & + \int \dd^{4}\mf x  \, F(\bm \xi) \delta\tau \tilde{\chi}'(\tau) \ell(\mf x).
\end{align}
Next, observe that to leading order we can write
\begin{equation}
    \tilde{\chi}(u) = \Theta(u - \tilde{\tau}_{i}) - \Theta(u - \tilde{\tau}_{f}) = \chi(u) + \delta \tau_{f} \delta(u - \tau_{f}) - \delta \tau_{i} \delta(u - \tau_{i}).
\end{equation}
Without loss of generality, we set $\tau_{i} = 0$. Thus, we obtain
\begin{equation}
    \tilde{\chi}'(u) = \delta(u - \tau_i) - \delta(u - \tau_{f}).
\end{equation}
The last term can be simplified as follows,
\begin{equation}
    \int \dd^{4} \mf x \, F(\bm \xi) \delta \tau (\delta(\tau - \tau_i) - \delta(\tau - \tau_{f})) \ell(\mf x) = -\delta \tau_{f} \int \dd^{3} \boldsymbol{\bm \xi} \, F(\bm \xi) \ell(\tau_{f}, \bm \xi)+\delta \tau_{i} \int \dd^{3} \boldsymbol{\bm \xi} \, F(\bm \xi) \ell(\tau_{i}, \bm \xi).
\end{equation}
Thus, to leading order in the perturbation of the trajectory followed by the smeared clock, the following holds:
\begin{align}
    \mathcal{T}(\tilde{C}_{\text{ideal}})& = \mathcal{T}(C_{\text{ideal}}) + \int \dd^{4} \mf x \, \delta \bm \xi \cdot \nabla F(\bm \xi) \chi(\tau)\ell(\mf x) - \int \dd^{4} \mf x\,  F(\bm \xi)\chi(\tau) \frac{g_{\mu \nu}u^{\mu}\delta u^{\nu}}{\ell(\mf x)} \nonumber \\ 
    & \qquad - \delta \tau_{f} \int \dd^{3} \boldsymbol{\bm \xi} \, F(\bm \xi) \ell(\tau_{f}, \bm \xi)+ \delta \tau_{i} \int \dd^{3} \boldsymbol{\bm \xi} \, F(\bm \xi) \ell(\tau_{i}, \bm \xi) + \mathcal{O}(\delta^2).
\end{align}
This expression can be simplified further in the FNC adapted to
$\mf z(\tau)$. In these coordinates,
$u^\mu=(1,\bm 0)$ and
$\delta u^\mu=
\bigl(\partial_\tau\delta\tau,
\partial_\tau\delta\bm\xi\bigr)$.
Since the FNC metric in flat spacetime is diagonal, it follows that
\begin{equation}
    g_{\mu\nu}u^\mu \delta u^\nu = g_{00}\partial_\tau (\delta \tau).   
\end{equation}
Finally, using $\ell(\mf x) = \sqrt{-g_{00}}$ in these coordinates, we conclude that 
\begin{equation}
    - g_{\mu\nu}u^\mu \delta u^\nu/\ell(\mf x) = \ell(\mf x) \partial_\tau(\delta \tau).
\end{equation}
Hence, 
\begin{align}
    \mathcal{T}(\tilde{C}_{\text{ideal}}) =& \mathcal{T}(C_{\text{ideal}}) + \int_0^{\tau_f} \dd \tau \int \dd^{3} \bm \xi \, \delta \bm \xi \cdot \nabla F(\bm \xi)\ell(\mf x) + \int_{\tau_i}^{\tau_f} \dd \tau \int \dd^{3} \bm \xi\,  F(\bm \xi) \ell(\mf x) \partial_\tau(\delta \tau) \nonumber \\ & - \delta \tau_{f} \int \dd^{3} \boldsymbol{\bm \xi} \, F(\bm \xi) \ell(\tau_{f}, \bm \xi)+ \delta \tau_{i} \int \dd^{3} \boldsymbol{\bm \xi} \, F(\bm \xi) \ell(\tau_{i}, \bm \xi) + \mathcal{O}(\delta^2).
\end{align}
Combining the last two terms by integration by parts yields the
following first-order relation between the times measured by the
original and perturbed smeared ideal clocks:
\begin{equation}
    \mathcal{T}(\tilde{C}_{\text{ideal}}) = \mathcal{T}(C_{\text{ideal}}) + \int_0^{\tau_f} \dd \tau \int \dd^{3} \bm \xi \, \delta \bm \xi \cdot \nabla F(\bm \xi)\ell(\mf x) - \int_{\tau_i}^{\tau_f} \dd \tau \int \dd^{3} \bm \xi\,  F(\bm \xi) \partial_\tau\ell(\mf x) \delta \tau + \mathcal{O}(\delta^2).
    \label{eq:perturbation_smeared_ideal_clock}
\end{equation}
Having established the general result for a smeared ideal clock, we
now turn to the QFT-based clock model introduced in the main text.
This clock consists of a finite-sized quantum system with internal
dynamics whose transition probability is modified by its interaction
with the vacuum field. It is specified by the triple
$C=\bigl(\mf z(\tau),F(\bm\xi),\Omega\bigr)$, where
$\mf z(\tau)$ is its reference trajectory, $F(\bm\xi)$ is its spatial
profile in the associated FNC, and $\Omega$ is the energy gap between
its excited and ground states. The time measured by the clock is
defined as
\begin{equation}
    \mathcal{T}(C) = \frac{\lambda^2}{\mathcal{F}}\int \dd V \dd V'F(\bm \xi) F(\bm \xi') \chi(\tau)\chi(\tau') e^{\ii \Omega(\tau - \tau')} W(\mf x, \mf x'),
    \label{eq:clock_def_appendix}
\end{equation}
where $W(\mf x,\mf x')$ is the Wightman function of the field and
$\mathcal{F}$ is the clock's decay rate in the long-time limit.
We now define
\begin{equation}
    \ell_{\Omega}(\mf x) \equiv  \frac{\lambda^2}{\mathcal{F}}\sqrt{-g_{00}}\Re\bigg(\int_{\tau_i}^{\tau_f} \dd \tau'\int \dd^3 \bm \xi'\sqrt{-g_{00}'} \, F(\bm \xi') e^{\ii \Omega(\tau(\mf x) - \tau')}W(\mf z_{\bm \xi}(\tau),\mf z_{\bm \xi'}(\tau'))\bigg),
    \label{eq:defLomega_appendix}
\end{equation}
with $g_{00}' = g_{00}(\tau', \bm \xi')$.
Then, $\mathcal{T}(C)$ can be cast in the same form as Eq.~\eqref{eq:ideal_smeared_appendix}, namely,
\begin{equation}
    \mathcal{T}(C) = \int_{\tau_i}^{\tau_f}\dd\tau \int \dd^3\bm \xi F(\bm \xi) \ell_\Omega(\mf x).
    \label{eq:QFT_clock_appendix}
\end{equation}
Indeed,
\begin{align}
    \int_{\tau_{i}}^{\tau_f}\dd\tau \int \dd^3\bm \xi F(\bm \xi) \ell_\Omega(\mf x)
     =& \frac{\lambda^2}{\mathcal{F}}\int_{\tau_i}^{\tau_f}\dd\tau \int \dd^3\bm \xi F(\bm \xi) \sqrt{-g_{00}} \Re\bigg(\int_{\tau_i}^{\tau_f} \dd \tau'\int \dd^3 \bm \xi'\sqrt{-g_{00}'} \, F(\bm \xi') e^{\ii \Omega(\tau(\mf x) - \tau')}W(\mf z_{\bm \xi}(\tau),\mf z_{\bm \xi'}(\tau'))\bigg) \nonumber \\ 
      =& \frac{\lambda^2}{\mathcal{F}}\int_{\tau_{i}}^{\tau_f}\dd\tau \int \dd^3\bm \xi \sqrt{-g_{00}} \, F(\bm \xi)\int_{\tau_i}^{\tau_f} \dd \tau'\int \dd^3 \bm \xi' F(\bm \xi')\sqrt{-g_{00}'} e^{\ii \Omega(\tau(\mf x) - \tau')}W(\mf z_{\bm \xi}(\tau),\mf z_{\bm \xi'}(\tau'))\nonumber\\
    =& \,\, \mathcal{T}(C),
\end{align}
where, in the last line, we used the fact that $\dd V  = \sqrt{-g_{00}}\, \dd \tau \dd^{3} \bm \xi$ in FNC in flat spacetime.

\subsection{The effective proper time density of QFT-based inertial clocks}

We now specialize the general construction to an inertial clock with
energy gap $\Omega$ and a spherically symmetric spatial profile
$F(\bm x)=F(|\bm x|)$. In inertial coordinates, the corresponding
effective proper-time density is
\begin{align}
    \ell_\Omega(\mf x) &= \frac{\lambda^2}{\mathcal{F}} \Re\left(\int_{\tau_i}^{\tau_f} \dd t' \int \dd^3 \bm x' F(|\bm x'|) e^{\ii \Omega(t-t')}W(t-t',\bm x - \bm x')\right) \nonumber \\
    &= \frac{\lambda^2}{\mathcal{F}} \frac{1}{(2\pi)^3}\Re\left(\int \dd^3 \bm k \int_{\tau_i}^{\tau_f} \dd t' \int \dd^3 \bm x' F(|\bm x'|) e^{\ii \Omega(t-t')} \frac{1}{2|\bm k|}e^{-\ii |\bm k| (t-t') + \ii \bm k \cdot (\bm x - \bm x')}\right) \nonumber \\
    &= \frac{\lambda^2}{\mathcal{F}} \frac{1}{(2\pi)^3}\Re\left(\int \dd^3 \bm k \frac{1}{2|\bm k|}\tilde{F}(|\bm k|) e^{\ii (\Omega - |\bm k|) t + \ii \bm k \cdot \bm x}\int_{\tau_i}^{\tau_f} \dd t' e^{-\ii (\Omega - |\bm k|) t'}\right),
\end{align}
where $\tilde{F}$ denotes the Fourier transform of $F$. Recall that,
for an inertial clock, the decay rate $\mathcal{F}$ is given by
Eq.~\eqref{eq:ideal_clock}. We now consider the long-time limit, in
which the $t'$ integration extends over the entire real line and,
in the distributional sense,
\begin{equation}
    \int_{-\infty}^{\infty}\dd t'\,
    e^{-\ii(\Omega-|\bm k|)t'}
    =
    2\pi\delta(\Omega-|\bm k|).
\end{equation}
It follows that
\begin{align}
    \ell_\Omega(\mf x)
    &=
    \frac{1}{2\pi \Omega \widetilde{F}(\Omega)^2}
    \Re\left[
    \int \dd^3 \bm k \,
    \frac{1}{2|\bm k|}
    \widetilde{F}(|\bm k|)
    e^{\ii(\Omega-|\bm k|)t+\ii \bm k\cdot \bm x}
    \delta(\Omega-|\bm k|)
    \right]
    \nonumber \\
    &=
    \frac{1}{2\Omega \widetilde{F}(\Omega)^2}
    \Re\left[
    \int_{0}^{\infty}\dd k
    \int_{0}^{\pi}\dd\theta\,
    k\sin\theta\,
    \widetilde{F}(k)
    e^{\ii(\Omega-k)t+\ii k|\bm x|\cos\theta}
    \delta(\Omega-k)
    \right]
    \nonumber \\
    &=
    \frac{1}{\Omega \widetilde{F}(\Omega)^2}
    \Re\left[
    \int_{0}^{\infty}\dd k\,
    k\widetilde{F}(k)
    e^{\ii(\Omega-k)t}
    \delta(\Omega-k)
    \operatorname{sinc}(k|\bm x|)
    \right]
    \nonumber \\
    &=
    \frac{1}{\Omega \widetilde{F}(\Omega)^2}
    \Omega \widetilde{F}(\Omega)\operatorname{sinc}(\Omega|\bm x|)=
    \frac{\operatorname{sinc}(\Omega|\bm x|)}{\widetilde{F}(\Omega)} .
\end{align}
When integrated in space against the spatial profile $F(\bm x)$, we obtain
\begin{align}
    \tilde{F}(|\bm k|) &= \int \dd^3 \bm x F(|\bm x|) e^{\ii \bm k \cdot \bm x} = 2\pi \int \dd |\bm x| \dd \theta \,\,|\bm x|^2 \sin\theta F(|\bm x|) e^{\ii |\bm k| |\bm x| \cos\theta} \\
    &=  4\pi \int \dd |\bm x| \, |\bm x|^2 F(|\bm x|) \text{sinc}(|\bm k| |\bm x|) = \int \dd^3 \bm x F(|\bm x|) \text{sinc}(|\bm k| |\bm x|).
\end{align}
Therefore,
\begin{equation}
    \int \dd^3\bm x\,
    F(\bm x)\ell_\Omega(\mf x)=1.
\end{equation}
Thus, although $\ell_\Omega(\mf x)$ may vary spatially, its average
over a spherically symmetric clock profile reproduces the classical
proper-time density. The clock is therefore insensitive to this
spatial dependence after averaging over its extent.
For a finite interaction interval symmetric about the origin,
$\tau_i=-T$ and $\tau_f=T$, let $k\equiv|\bm k|$. The effective
proper-time density is then
\begin{align}
    \ell_\Omega(\mf x)
    &=
    \frac{1}{\Omega\tilde{F}(\Omega)^2}
    \Re\left[
        \int_0^\infty\dd k\,
        k\tilde{F}(k)
        e^{\ii(\Omega-k)t}
        \frac{T}{\pi}
        \operatorname{sinc}\bigl((\Omega-k)T\bigr)
        \operatorname{sinc}(k|\bm x|)
    \right]
    \\
    &=
    \frac{1}{\Omega\tilde{F}(\Omega)^2}
    \int_0^\infty\dd k\,
    k\tilde{F}(k)
    \cos\bigl((\Omega-k)t\bigr)
    \frac{T}{\pi}
    \operatorname{sinc}\bigl((\Omega-k)T\bigr)
    \operatorname{sinc}(k|\bm x|).
\end{align}
The preceding long-time result is recovered using the distributional
limit
\begin{equation}
    \lim_{T\to\infty}
    \frac{T}{\pi}
    \operatorname{sinc}\bigl((\Omega-k)T\bigr)
    =
    \delta(\Omega-k).
\end{equation}

\subsection {Observer-dependent time in the QFT clock model}

Consider now a clock
$\tilde C=(\tilde{\mf z}(\tilde\tau),
F(\tilde{\bm\xi}),\Omega)$ whose reference trajectory is related to
that of $C$ by
\begin{equation}
    \tilde{\mf z}=\mf z+\delta\mf z.
\end{equation}
We assume that the two reference trajectories connect the same pair
of events, $\mf x_1$ and $\mf x_2$, so that $\delta\mf z$ vanishes at
both endpoints. Let $(\tau,\bm\xi)$ denote the Fermi normal
coordinates associated with the unperturbed trajectory
$\mf z(\tau)$, and let $(\tilde\tau,\tilde{\bm\xi})$ denote those
associated with the perturbed trajectory. As shown in
Sec.~\ref{sec:FNC_perturbation_justified}, their relation to first
order in the perturbation is
\begin{equation}
    \tilde \tau=\tau+\delta\tau,
    \qquad
    \tilde{\bm \xi}=\bm \xi+\delta\bm \xi.
\end{equation}
 To first order in the perturbation, we have
\begin{align}
    \mathcal{T}(\tilde C)
    =&\,\frac{\lambda^2}{\mathcal{F}}
    \int \dd V\,\dd V'\,
    F(\tilde{\bm \xi})F(\tilde{\bm \xi}')
    \tilde\chi(\tilde\tau)
    \tilde\chi(\tilde\tau')
    e^{\ii\Omega(\tilde\tau-\tilde\tau')}
    W(\mf x,\mf x')
    \nonumber\\
    =&\,\frac{\lambda^2}{\mathcal{F}}
    \int \dd V\,\dd V'\,
    F(\bm \xi+\delta\bm \xi)
    F(\bm \xi'+\delta\bm \xi')
    \tilde\chi(\tau+\delta\tau)
    \tilde\chi(\tau'+\delta\tau')
    e^{\ii\Omega(\tau-\tau')}
    e^{\ii\Omega(\delta\tau-\delta\tau')}
    W(\mf x,\mf x')
    \nonumber\\
    =&\,\mathcal{T}(C)
    +\frac{\lambda^2}{\mathcal{F}}
    \int \dd V\,\dd V'\,
    \bigl[
        \delta\bm \xi\cdot\nabla F(\bm \xi)F(\bm \xi')
        +
        F(\bm \xi)\delta\bm \xi'\cdot\nabla F(\bm \xi')
    \bigr]
    \chi(\tau)\chi(\tau')
    e^{\ii\Omega(\tau-\tau')}
    W(\mf x,\mf x')
    \nonumber\\
    &+\frac{\lambda^2}{\mathcal{F}}
    \int \dd V\,\dd V'\,
    F(\bm \xi)F(\bm \xi')
    \bigl[
        \delta\tau\,\chi'(\tau)\chi(\tau')
        +
        \chi(\tau)\delta\tau'\,\chi'(\tau')
    \bigr]
    e^{\ii\Omega(\tau-\tau')}
    W(\mf x,\mf x')
    \nonumber\\
    &+\frac{\lambda^2\ii\Omega}{\mathcal{F}}
    \int \dd V\,\dd V'\,
    F(\bm \xi)F(\bm \xi')
    \chi(\tau)\chi(\tau')
    (\delta\tau-\delta\tau')
    e^{\ii\Omega(\tau-\tau')}
    W(\mf x,\mf x')
    +\mathcal{O}(\delta^{2}) .
\label{eq:T_shifted_expansion_inertial}
\end{align}
The switching function of the perturbed clock is
\begin{align}
    \tilde{\chi}(u)
    &=
    \Theta(u-\tilde{\tau}_i)
    -
    \Theta(u-\tilde{\tau}_f)
    \nonumber\\
    &=
    \Theta(u-\tau_i-\delta\tau_i)
    -
    \Theta(u-\tau_f-\delta\tau_f)
    \nonumber\\
    &=
    \Theta(u-\tau_i)
    -
    \Theta(u-\tau_f)
    -
    \delta\tau_i\,\delta(u-\tau_i)
    +
    \delta\tau_f\,\delta(u-\tau_f)
    +
    \mathcal{O}(\delta^2)
    \nonumber\\
    &=
    \chi(u)
    -
    \delta\tau_i\,\delta(u-\tau_i)
    +
    \delta\tau_f\,\delta(u-\tau_f)
    +
    \mathcal{O}(\delta^2).
    \label{eq:shifted_switching_expansion}
\end{align}
The derivative of the unperturbed switching function is
\begin{equation}
    \chi'(u)
    =
    \delta(u-\tau_i)
    -
    \delta(u-\tau_f),
\end{equation}
and, to the order considered here,
$\tilde{\chi}'(u)=\chi'(u)+\mathcal{O}(\delta)$.

Substituting Eq.~\eqref{eq:shifted_switching_expansion} into
Eq.~\eqref{eq:T_shifted_expansion_inertial}, the endpoint terms
arising from the displacement of the switching support cancel those
generated by the expansion of
$\tilde{\chi}(\tau+\delta\tau)$. This cancellation follows from the
assumption that the two clocks connect the same endpoint events.
Expressing the invariant volume element in the FNC associated with
the unperturbed trajectory as $\dd V
    =
    \sqrt{-g_{00}(\tau,\bm\xi)}\,
    \dd\tau\,\dd^3\bm\xi$
we obtain
\begin{align}
    \mathcal{T}(\tilde C)
    =&\,\mathcal{T}(C)
    \nonumber\\
    &+\frac{\lambda^2}{\mathcal{F}}
    \int_{\tau_i}^{\tau_f}\dd\tau
    \int_{\tau_i}^{\tau_f}\dd\tau'
    \int \dd^{3}\bm \xi  \,\dd^{3}\bm \xi' \sqrt{-g_{00}}\sqrt{-g_{00}'}\,
    \bigl[
        \delta\bm \xi\cdot\nabla F(\bm \xi)F(\bm \xi')
        +
        F(\bm \xi)\delta\bm \xi'\cdot\nabla F(\bm \xi')
    \bigr]
    e^{\ii\Omega(\tau-\tau')}
    W(\tau,\bm \xi;\tau',\bm \xi')
    \nonumber\\
    &+\frac{\lambda^2\ii\Omega}{\mathcal{F}}
    \int_{\tau_i}^{\tau_f}\dd\tau
    \int_{\tau_i}^{\tau_f}\dd\tau'
    \int \dd^{3}\bm \xi\,\dd^{3}\bm \xi'\,     \sqrt{-g_{00}}\sqrt{-g_{00}'}
    F(\bm \xi)F(\bm \xi')
    (\delta\tau-\delta\tau')
    e^{\ii\Omega(\tau-\tau')}
    W(\tau,\bm \xi;\tau',\bm \xi')
    +\mathcal{O}(\delta^{2}) .
    \label{eq:T_shifted_after_endpoint_cancellation}
\end{align}
Using $W(\tau,\bm \xi;\tau',\bm \xi') = W(\tau',\bm \xi';\tau,\bm \xi)^{*}$, the primed and unprimed contributions can be combined. This gives
\begin{align}
    \mathcal{T}(\tilde C)
    =&\,\mathcal{T}(C)
    \nonumber\\
    &+\frac{2\lambda^2}{\mathcal{F}}\Re
    \bigg[
    \int_{\tau_i}^{\tau_f}\dd\tau
    \int_{\tau_i}^{\tau_f}\dd\tau'
    \int \dd^{3}\bm \xi\,\dd^{3}\bm \xi'\,    \sqrt{-g_{00}}\sqrt{-g_{00}'}
    \delta\bm \xi\cdot\nabla F(\bm \xi)F(\bm \xi')
    e^{\ii\Omega(\tau-\tau')}
    W(\tau,\bm \xi;\tau',\bm \xi')
    \bigg]
    \nonumber\\
    &+\frac{2\lambda^2}{\mathcal{F}}\Re
    \bigg[
    \int_{\tau_i}^{\tau_f}\dd\tau
    \int_{\tau_i}^{\tau_f}\dd\tau'
    \int \dd^{3}\bm \xi\,\dd^{3}\bm \xi'\,    \sqrt{-g_{00}}\sqrt{-g_{00}'}
    F(\bm \xi)F(\bm \xi')
    \delta\tau\,
    \partial_{\tau}
    \bigl(e^{\ii\Omega(\tau-\tau')}\bigr)
    W(\tau,\bm \xi;\tau',\bm \xi')
    \bigg]
    +\mathcal{O}(\delta^{2}) .
    \label{eq:T_shifted_combined_real}
\end{align}
The last term can be reorganized using
\begin{equation}
    \partial_{\tau}
    \bigl(e^{\ii\Omega(\tau-\tau')}\bigr)
    W(\tau,\bm \xi;\tau',\bm \xi')
    =
    \partial_{\tau}
    \bigl[
        e^{\ii\Omega(\tau-\tau')}
        W(\tau,\bm \xi;\tau',\bm \xi')
    \bigr]
    -
    e^{\ii\Omega(\tau-\tau')}
    \partial_{\tau}
    W(\tau,\bm \xi;\tau',\bm \xi') .
\end{equation}
Moreover,
\begin{equation}
    \partial_{\tau}
    W(\tau,\bm \xi;\tau',\bm \xi')
    =
    \partial_{\tau}
    W\bigl(\mf z_{\bm \xi}(\tau),\mf z_{\bm \xi'}(\tau')\bigr)
    =
    \dv{z_{\bm \xi}^{\mu}}{\tau}
    \partial_{\mu}
    W\bigl(\mf z_{\bm \xi}(\tau),\mf z_{\bm \xi'}(\tau')\bigr).
\end{equation}
We now use the definition of $\ell_{\Omega}(\mf x)$ as in Eq.~\eqref{eq:defLomega_appendix}, together with
\begin{equation}
    \ell_{\Omega,\mu}(\mf x)
    \equiv
    \frac{\lambda^2}{\mathcal{F}}
    \sqrt{-g_{00}}\,
    \Re
    \bigg[
    \int_{\tau_i}^{\tau_f}\dd\tau'
    \int \dd^{3}\bm \xi'\, \sqrt{-g_{00}'}
    F(\bm \xi')
    e^{\ii\Omega(\tau(\mf x)-\tau')}
    \partial_{\mu}
    W\bigl(\mf z_{\bm \xi}(\tau),\mf z_{\bm \xi'}(\tau')\bigr)
    \bigg].
    \label{eq:def_lOmega_mu_polished}
\end{equation}
Since $\sqrt{-g_{00}}$ may depend on $\tau$ for a general accelerated Fermi frame, we also introduce the shorthand
\begin{equation}
    \dot{\ell}_{\Omega}(\mf x)
    \equiv
    \partial_{\tau}\ell_{\Omega}(\mf x)
    -
    \ell_{\Omega}(\mf x)\,
    \partial_{\tau}\log(\sqrt{-g_{00}}) .
    \label{eq:def_kernel_tau_derivative}
\end{equation}
Then, combining the previous expressions, we finally obtain
\begin{align}
    \mathcal{T}(\tilde C)
    & =\mathcal{T}(C)
    +
    2
    \int_{\tau_i}^{\tau_f}\dd\tau
    \int \dd^{3}\bm \xi\,
    \delta\bm \xi\cdot\nabla F(\bm \xi)\,
    \ell_{\Omega}(\mf x)+
    2
    \int_{\tau_i}^{\tau_f}\dd\tau
    \int \dd^{3}\bm \xi\,
    F(\bm \xi)\,
    \delta\tau\,
    \dot{\ell}_{\Omega}(\mf x) \nonumber \\ & \qquad -
    2
    \int_{\tau_i}^{\tau_f}\dd\tau
    \int \dd^{3}\bm \xi\,
    F(\bm \xi)\,
    \delta\tau\,
    \ell_{\Omega,\mu}(\mf x)
    \dv{z_{\bm \xi}^{\mu}}{\tau}
    +\mathcal{O}(\delta^{2}) .
    \label{eq:T_shifted_general}
\end{align}
If $\sqrt{-g_{00}}$ is independent of $\tau$, as in inertial motion
or uniform acceleration with constant tetrad components $a_i$, then
$\partial_\tau\log\sqrt{-g_{00}}=0$ and
$\dot{\ell}_\Omega(\mf x)=\partial_\tau\ell_\Omega(\mf x)$.
Equation~\eqref{eq:T_shifted_general} therefore reduces to
\begin{align}
    \mathcal{T}(\tilde C)
    =&\,\mathcal{T}(C)
    +
    2
    \int_{\tau_i}^{\tau_f}\dd\tau
    \int \dd^{3}\bm \xi\,
    \delta\bm \xi\cdot\nabla F(\bm \xi)\,
    \ell_{\Omega}(\mf x)+
    2
    \int_{\tau_i}^{\tau_f}\dd\tau
    \int \dd^{3}\bm \xi\,
    F(\bm \xi)\,
    \delta\tau\,
    \partial_{\tau}\ell_{\Omega}(\mf x)
    \nonumber\\
    &-
    2
    \int_{\tau_i}^{\tau_f}\dd\tau
    \int \dd^{3}\bm \xi\,
    F(\bm \xi)\,
    \delta\tau\,
    \ell_{\Omega,\mu}(\mf x)
    \dv{z_{\bm \xi}^{\mu}}{\tau}
    +\mathcal{O}(\delta^{2}) .
    \label{eq:T_shifhter_particular}
\end{align}
Equation~\eqref{eq:T_shifted_general} is the QFT analogue of the
variation obtained for the smeared ideal clock in
Eq.~\eqref{eq:perturbation_smeared_ideal_clock}. It gives the
leading-order change in the observer-dependent time measured by the
QFT-based clock. The first correction has the same origin as its
classical counterpart: it results from the deformation of the clock's
spatial profile within its world tube. In the QFT model, this
contribution carries an additional factor of two because the spatial
profile appears at both arguments of the Wightman function.
The second correction is also analogous to the ideal-clock result,
although its sign and numerical factor differ and the geometric
proper-time density $\ell(\mf x)$ is replaced by the effective density
$\ell_\Omega(\mf x)$. The final correction, which is proportional to
the four-velocity of the trajectories $z_{\bm\xi}$ within the clock's
world tube, has no counterpart in the smeared ideal case. This
term arises from the variation of the Wightman function along the
world tube and accounts for the change in how the perturbed clock
samples vacuum correlations.

\twocolumngrid

\bibliography{references}

\end{document}